\documentclass[11pt,a4paper]{emulateapj}

\usepackage{natbib}
\usepackage{epsfig}
\usepackage{amsmath}
\usepackage{longtable, booktabs}
\usepackage{url}
\usepackage{float}

\shorttitle{Selecting Sagittarius}
\shortauthors{E.A. Hyde et al.}

\begin{document}

\title{Selecting Sagittarius: \\
    Identification and Chemical Characterization\\ of the Sagittarius Stream}

\author{E.A. Hyde\altaffilmark{1}}
\affil{University of Western Sydney, Locked Bag 1797, Penrith South DC, NSW 1797, Australia}
\email{E.Hyde@uws.edu.au}
\altaffiltext{1}{Macquarie University, Physics and Astronomy, NSW 2109, Australia}
\author{S. Keller}
\affil{Research School of Astronomy and Astrophysics, Australian National University, Canberra, ACT 2601, Australia}
\author{D.B. Zucker\altaffilmark{2}}
\affil{Macquarie University, Physics and Astronomy, NSW 2109, Australia}
\altaffiltext{2}{Australian Astronomical Observatory, PO Box 296 Epping, NSW 1710, Australia}
\author{R. Ibata, and A. Siebert}
\affil{Observatoire astronomique de Strasbourg, Universit\'e de Strasbourg, CNRS, UMR 7550, 11 rue de l'Universit\'e, F-67000 Strasbourg, France}
\author{G.F. Lewis}
\affil{Sydney Institute for Astronomy, School of Physics, The University of Sydney, NSW 2006, Australia}
\author{J. Penarrubia}
\affil{ROE, The University of  Edinburgh, Institute for Astronomy, Edinburgh EH9 3HJ, UK}
\author{M. Irwin and G. Gilmore}
\affil{Institute of Astronomy, University of Cambridge, Madingley Road, Cambridge CB3 0HA, UK}
\author{R.R. Lane}
\affil{Departamento de Astronom\'ia Universidad de Concepci\'on, Casilla 160 C, Concepci\'on, Chile}
\author{A. Koch}
\affil{Landessternwarte, Zentrum f\"ur Astronomie der Universit\"at Heidelberg,
K\"onigstuhl 12, 69117 Heidelberg, Germany}
\author{A.R. Conn}
\affil{Sydney Institute for Astronomy, School of Physics, The University of Sydney, NSW 2006, Australia}
\author{F.I. Diakogiannis}
\affil{International Center for Radio Astronomy Research, University of Western Australia, 35 Stirling Highway, Crawley, WA 6009, Australia}
\author{S. Martell}
\affil{Department of Astrophysics, School of Physics, University of New South Wales, Sydney, NSW 2052, Australia}

\begin{abstract}
Wrapping around the Milky Way, the Sagittarius stream is the dominant substructure in the halo. Our statistical selection method has allowed us to identify 106 highly likely members of the Sagittarius stream.  Spectroscopic analysis of metallicity and kinematics of all members provides us with a new mapping of the Sagittarius stream. We find correspondence between the velocity distribution of stream stars and those computed for a triaxial model of the Milky Way dark matter halo. The Sagittarius trailing arm exhibits a metallicity gradient, ranging from $-0.59$ dex to $-0.97$  dex over 142$^{\circ}$. This is consistent with the scenario of tidal disruption from a progenitor dwarf galaxy that possessed an internal metallicity gradient. We note high metallicity dispersion in the leading arm, causing a lack of detectable gradient and possibly indicating orbital phase mixing. We additionally report on a potential detection of the Sextans dwarf spheroidal in our data.
  
\end{abstract}


\section{Introduction}
\label{sec:level1}

\begin{figure*}[b!t]
\includegraphics[width=0.98\textwidth]{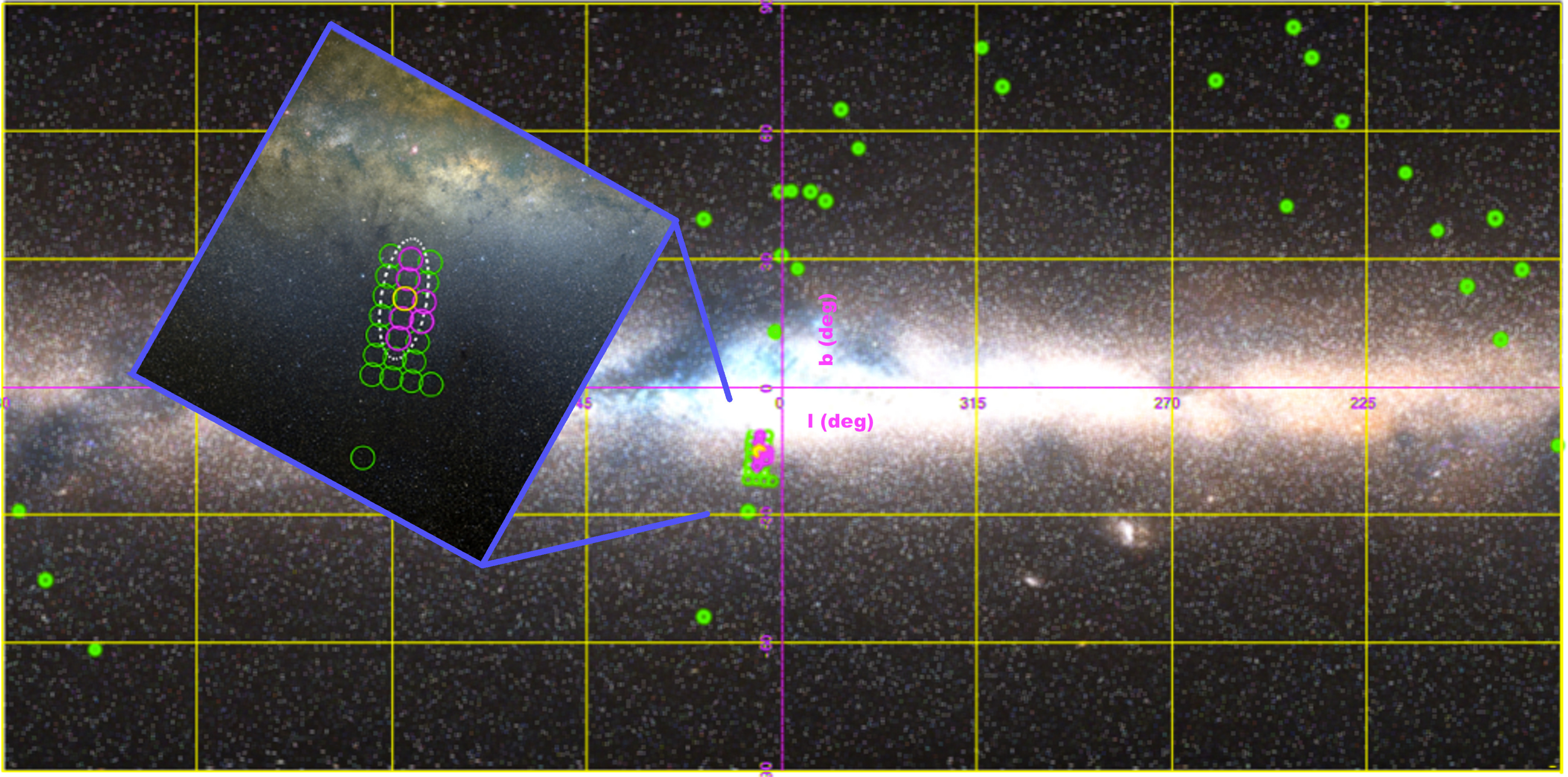}
\caption{Sgr dwarf and stream regions observed are shown as green  circles; each observation is 2 degrees in diameter. The Sgr dwarf core is located in the center yellow circle and the central rectangular block makes up the `central region' F01.  The magenta circles are the first observations taken, used to show no rotation in the core of Sagittarius \citep{2011ApJ...727L...2P}. Background image constructed from online data \citep{2009PASP..121.1180M} and uses a Mercator type projection. We show the coordinate grid for Galactic l, b, positions where zero marks the center of the Milky Way.}
\label{fig:core}
 \end{figure*}

The Sagittarius (Sgr) dwarf was discovered by \cite{1994Natur.370..194I} and its associated stream was mapped using M-giant stars in the Two Micron All Sky Survey \citep[2MASS;][]{2006AJ....131.1163S} catalog by \cite{2003ApJ...599.1082M}. This stream traces an arc which wraps at least twice around our Milky Way galaxy, passing through the gravitational potential of the Milky Way. This stream can help us probe the dark matter halo around the Milky Way \citep{2001ApJ...551..294I}, as well as understand the progenitor of the current dwarf and stream system. In particular, the leading arm of Sgr is known to be sensitive both kinematically and spatially to the shape of the dark matter halo  \citep[e.g.,][]{2002MNRAS.332..921I,2004AJ....128..245M,2005AJ....129..189V,2009ApJ...703L..67L}.

The Sgr stream was kinematically traced in K-giants by \cite{2012AJ....143...88C}. They found that the Milky Way triaxial dark matter halo model of \cite{2010ApJ...714..229L} allowed them to improve their match between modeling and data for the Sgr stream. The triaxial model has since gained further support with the data from blue horizontal branch stars in the  Sgr stream \citep{2009ApJ...700.1282Y}, where velocities closely match those predicted in the triaxial model case.  Although several attempts have been made to map the stream \citep[for example,][]{2003ApJ...599.1082M,2012ApJ...760...95J,2014MNRAS.437..116B,2014A&A...564A..18P} there are still conflicts on Milky Way halo shapes determined from Sgr models \citep[e.g.,][]{2004ApJ...610L..97H,2005ApJ...619..800J,2010ApJ...714..229L,2010MNRAS.406..922T,2013MNRAS.428..912D,2013ApJ...773L...4V,2013ApJ...765L..15I}).

One particular feature that models have had some issue reproducing in the Sgr stream is an apparent bifurcation discovered by \cite{2006ApJ...642L.137B}. This bifurcated nature of the stream has been confirmed in both Galactic hemispheres \citep{2012ApJ...750...80K,2012arXiv1211.2817S} and was proposed to result from a spherical dark matter halo by \cite{2006ApJ...651..167F}.  It was shown by \cite{2010MNRAS.408L..26P} that a bifurcation can naturally arrive from models using a disky progenitor for Sgr, but \cite{2011ApJ...727L...2P} found none of the internal rotation that would be indicative of such a scenario. \cite{2013AJ....145..163N} have suggested that the two tails of the Sgr stream may have different progenitors. An additional possibility explored by \cite{2012AJ....143...88C} is that the bifurcation may not need to be fit to current models if it was caused by some dramatic recent event.  However, recent tracing of several Sgr stream members by \cite{2014MNRAS.437..116B} led them to conclude that no existing simulation for disruption could explain the observational data  and the problem has yet to be solved. 





 Simulations by \cite{2005ApJ...627..647B} prefer triaxial dark matter haloes for galaxies when producing large scale structure formations but it may not necessarily be required in the case of the Sgr dwarf \citep{2013ApJ...765L..15I}. Although there is some evidence for a Milky Way triaxial dark matter halo model, there are still numerous questions to be answered about the progenitor of the stream, in particular its metallicity.  Previous mapping of the core and stream of Sgr have included estimates of the metallicity distribution function (MDF). However, areas which show a mixed orbital phase and the gradients in metallicity with distance along the stream remain unconvincingly mapped. The mixed orbital phase regions occur when debris or stars stripped from the satellite orbit land within a range of azimuthal time periods about the satellite's own. This scenario would  lead to phase mixing of debris ahead and behind the satellite along its orbit, and \cite{1998ApJ...495..297J}  predicts that the tidal streamers from the Sagittarius dwarf galaxy currently extend over more than 2$\pi$ in azimuth along their orbits. The range extended by the debris at different points would increase the type or age of stars present at a given point in the stream. For this reason some areas of the stream are predicted to have a wider range in metallicity values than others \citep{2005ApJ...619..807L}.

 The population near the core of Sgr has been found to have a mean value of [Fe/H] $\sim -0.5$ dex \citep{2002IAUS..207..168C,2004A&A...414..503B,2005A&A...441..141M}.  This value is known to vary as we move out from the core, potentially reflecting the properties of the original progenitor of the dwarf and stream.

In the leading arm of the Sgr stream, \cite{2007ApJ...670..346C} found a mean metallicity of $\langle[$Fe/H$]\rangle = -0.72$ dex.  \cite{2012AJ....143...88C} find K-giant members of Sgr which are notably more metal-poor ($\langle[$Fe/H$]\rangle = -1.7 \pm 0.3$ dex). However, a wide range of metallicities throughout the stream are identified by both \cite{2007ApJ...670..346C} and \cite{2012AJ....143...88C}. This broad range of metallicities supports the theory of \cite{2004ApJ...601..242M}, who suggest that the progenitor to the Sgr dwarf and stream shed its layers (over which there must have been an intrinsic MDF gradient) successively in its orbit around the Milky Way. This could indicate a mixed orbital phase in the leading arm as described by the model of  \cite{2005ApJ...619..807L}. 

The core region of Sgr is predicted to be more metal rich than stars in the stream as the stream stars would be represented by debris lost some 3.5 orbits ($\sim2.5-3$ Gyr) ago  \citep[e.g.,][]{2005ApJ...619..807L,2009ApJ...703L..67L,2010ApJ...714..229L}.  Though the mean metallicity over the leading arm has been found to decrease with distance from the core by \cite{2007ApJ...670..346C}, the trailing arm was not mapped, and the wide range of metallicities present in the data prevented a strong detection of a gradient. \cite{2010ApJ...720..940K} show a significant gradient in the stream at $-(2.4 \pm 0.3)\times 10^{-3}$ dex/degree $(-(9.4 \pm 1.1)\times 10^{-4}$ dex/kpc) from the main body of Sgr.  This is interpreted by \cite{2010ApJ...720..940K} as indicative of a similar gradient in the progenitor of the Sgr dwarf and stream system.  

\subsection{Target Selection}
In this paper, we present a new selection of probable Sgr member stars in the stream using a statistical selection method, and show their metallicity distribution. Individual star velocities of the stream will be discussed in future stream analysis paper (R. Ibata et al. in prep). Our mapping of the Sgr stream includes both leading and trailing arm stars as shown in Figure \ref{fig:core}. We examine phase mixed areas of the stream as well as gradients in metallicity along the stream (away from the core). We include information for known features (overdensities and streams) that are near the Sgr stream at various points on the sky, namely the Sextans Dwarf  \citep[e.g.,][]{1990MNRAS.244P..16I}, the Virgo Overdensity \citep[e.g.,][]{2010PASA...27...45K} and the globular cluster Palomar 5 \citep[e.g.,][]{2002AJ....124.1497O} for comparison.

 A brief overview of the data reduction and analysis is outlined in Sections \ref{sec:data} and \ref{sec:analis}, and the selection of stream candidates is given in Section \ref{sec:sgrsel}. The core sample (the center region shown in the outlined box of Figure \ref{fig:core}) is hereafter described as `F01', the remaining stream fields shown are divided into 29 regions forming F02 to F30 as shown later in Section \ref{sec:mdisstream}.

In Section \ref{sec:compmod} we compare our candidate Sgr star properties to points rendered by models of the stream. The resulting velocity and metallicity distributions are then presented in Section \ref{sec:dis} and compared to the model output from a triaxial halo. We discuss a possible detection of the Sextans dwarf in Section \ref{sec:over} and summarize our conclusions in Section \ref{sec:conc}.

Candidate Sgr member stars were selected using two different methods for the two main observing programs that were carried out.  Firstly we selected stars in the stream as part of a joint observing campaign, and later the main core area was observed as part of a second, more detailed campaign to look for evidence of rotation in the core.

For stars in the stream we based our selection on photometry from the 2MASS catalog, supplemented (where applicable) with optical photometry from the Canada-France Hawaii Telescope (CFHT). Candidate Sgr stream stars  with 2MASS photometry were selected to reside within the color and magnitude confines of the following box: $J-K$ = (0.95, 1.35, 1.10, 0.95), $K$ = (13.0, 10.0, 10.0, 11.125). Targets which possessed CFHT photometry were selected as potential members of the following broad evolutionary phases: blue horizontal branch (BHB), main-sequence (MS), red clump (RC), and red giant branch (RGB) stars. The resulting pointings contained stars spanning the magnitude range $12 < V < 18$ (corresponding to a g$=20$ limit). 

 The second main observing campaign that was carried out was on the core region of the Sgr dwarf.  For this campaign we observed stars within about $\sim6^{\circ}$ of the center of the Sgr dwarf. The targets were selected based on the following criteria for extinction corrected J and K magnitudes from 2MASS:  $9 < K < 13$ and $J-K > (20 - K)/90$, this gives a resulting range in brightness between $12 < V < 15$. This selection corresponds to the region chosen by \cite{2004AJ....128..245M}, and is known to display a clear trend in velocities across orbital longitude.

\section{Observations and Data Reduction}
\label{sec:data}

Once selected, candidates were subsequently targeted with the AAOmega~\citep{2006SPIE.6269E..14S} fibre-fed multi-object spectrograph on the 3.9m Anglo-Australian Telescope (AAT). We utilize the 1500V and 1700D gratings of the AAOmega instrument with resolutions of R$= 8000$ and $10000$ in the blue and red respectively (the wavelength range is 4250-6000 $\mathrm{\AA}$ in the blue and 8450-9000 $\mathrm{\AA}$ in the red)\citep{2006SPIE.6269E..14S}. This enables us to target the spectral regions around the Mg I 5170 and Ca II 8500 $\mathrm{\AA}$ triplets. The AAOmega spectrograph possesses 392 fibers available for targets in a single pointing. All pointings had at least 25 fibers assigned to sky positions. Target stars were observed as allowed using the \emph{configure} program to assign fibers to stars. The observations are shown in Figure \ref{fig:core}. The data for both the stream and core observing programs were obtained on several observing runs spanning 2009-2011. 

The spectral reduction was undertaken using the \emph{2dfdr}\footnote{\texttt{http://www.aao.gov.au/AAO/2df/aaomega}} data reduction program of the Australian Astronomical Observatory \citep{AAO2dfdr}.  This performs bias correction, flat-fielding and optimal fiber extraction. Arc lamp exposures allowed for wavelength calibration between each set of observations. Typically, a pointing was observed in sets of three exposures, each with an exposure time of 1200 seconds. Cosmic ray reduction was then achieved by median combining the multiple object frames. We measure an elimination of $\sim$90\% of the contribution for all sky lines in the \emph{2dfdr} reduction. 

We use a modified version of the Radial Velocity Experiment (RAVE)  \citep{2006AJ....132.1645S,2008AJ....136..421Z,2011AJ....141..187S} pipeline to determine metallicity, surface gravity ($\mathrm{\log(g)}$), and effective temperature ($\mathrm{T_{eff}}$) for our target stars.  We estimate the velocities from a template fit on the Ca II 8500 $\mathrm{\AA}$ triplet using the IRAF\footnote{The IRAF software package is distributed by the National Optical Astronomy Observatories, which are operated by the Association of Universities for Research in Astronomy, Inc., under cooperative agreement with the National Science Foundation.} task \emph{fxcor}, which we then converted into heliocentric values ($\mathrm{V_{hel}}$) using the IRAF task \emph{rvcorrect}.  

After the heliocentric velocity conversion we transform the spatial coordinates into Galactic $l$, $b$ and convert the heliocentric radial velocity into a Galactocentric radial velocity ($\mathrm{V_{GSR}}$) using:
\begin{equation}\label{eqn:v}
\begin{aligned}
\mathrm{V_{GSR} = V_{hel}+9\cos(l)\cos(b) + 232\sin(l)\cos(b) +7\sin(b)}
\end{aligned}
\end{equation}
which has units of km s$^{-1}$ as described in \cite{1999A&A...341..437B}.

\section{Analysis}
\label{sec:analis}

 \begin{figure}[htb]
\begin{center}
\includegraphics[width=0.5\textwidth]{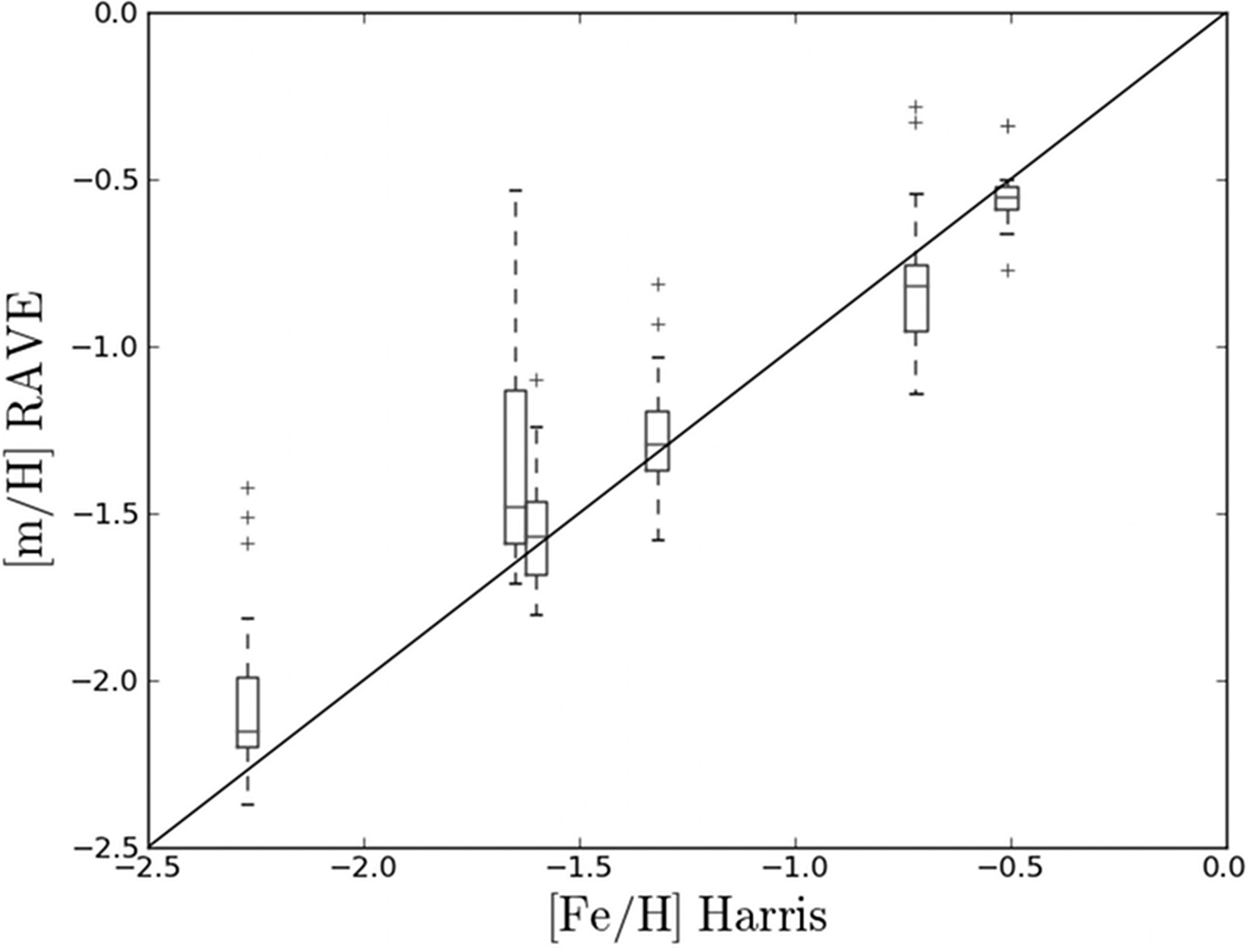}
\end{center}
\caption{ The $[$m/H$]$ calibration using six reduced calibration clusters. The solid line shows one to one correspondence. Each box corresponds to one of the clusters given in Table \ref{tab:table1}. The horizontal line in each box is the median for  the data measured on that cluster, the 25th percentile is given by the upper edge of the box, the 75th percentile is given by the lower edge of the box and the maximum and minimum are at the ends of the whiskers. The plus symbols give the N sigma outliers in the distribution.}
\label{fig:whiscal}
 \end{figure}

The modified RAVE pipeline metallicities  are total metallicities, referred to hereafter as [m/H]. The abundances are derived from analysis of the wavelength region around the Ca II 8500 $\mathrm{\AA}$ triplet using a library of spectral templates.  Once an input heliocentric correction is given for the observation in question the RAVE pipeline derives a temperature estimate and then iteratively determines the best spectral template match\footnote{ A. Siebert, private communication.}. The weights of the best match are determined by a $\chi^{2}$ routine described in \cite{2008AJ....136..421Z}. Conservative estimates of errors for a spectrum with an average signal-to-noise ratio (S/N $\sim 40$) are 400 K in temperature, 0.5 dex in gravity, and 0.2 dex in metallicity \citep{2008AJ....136..421Z}.

 In Figure \ref{fig:whiscal} we compare $[$m/H$]$ metallicity values from the modified RAVE pipeline with the literature values for several stellar clusters: 47 Tucanae, NGC 288, M30, M2, Melotte 66, and NGC 1904  from the Harris Catalogue \citep[2010 edition;][]{1996AJ....112.1487H}, as listed in Table \ref{tab:table1}.  The clusters were observed with the  AAT using the same instrument settings as the Sgr stream targets. The observations of these clusters were reduced and processed in the same way as the Sgr data. The results of the comparison between the literature metallicities and those obtained with the modified RAVE pipeline are shown in Figure~\ref{fig:whiscal}. The difference between the calculated median shown in Figure \ref{fig:whiscal} and the literature value of the mean metallicity for the cluster is given by $\mathrm{\Delta_{m} = [Fe/H]_{lit} - [m/H]_{mean}}$, as shown in Table \ref{tab:table1}. In all cases the difference $\mathrm{\Delta_{m} \leq 0.31}$ dex.  As four of the six clusters in Table \ref{tab:table1} have a mean greater than the median, there is a tendency in this data towards a positive skew (greater metallicity values).  Due to this skew there is closer correspondence between the median $\mathrm{[m/H]_{med}}$ and the literature mean value $\mathrm{[Fe/H]_{lit}}$. We treat $[$m/H$]$ as equivalent to [Fe/H] and will use the [m/H] as determined by the modified RAVE pipeline as metallicity for the rest of this paper.

\begin{table}[htb]
\begin{center}
\caption{The Calibration Clusters.\label{tab:table1}}
\begin{tabular}{cccccccc}
Cluster &Num & $\mathrm{[Fe/H]_{lit}}$ &$\mathrm{[m/H]_{med}}$ &$\mathrm{[m/H]_{mean}}$ &$\mathrm{\Delta_{m}}$\\
 &&dex & dex &dex &dex\\
\tableline
47 Tucanae	&67    	&-0.72	&-0.82	&-0.83	&0.11 	\\
NGC 288   	&39    	&-1.32 	&-1.29	&-1.27	&-0.05	\\
M 30			&25    	&-2.27 	&-2.15	&-2.06	&-0.21	\\
M 2			&17    	&-1.65 	&-1.48	&-1.34	&-0.31	\\
Melotte 66	&17 		&-0.51	&-0.55	&-0.55	&-0.04$^{6}$	\\
NGC 1904	&22  		&-1.60	&-1.57	&-1.54	&-0.06	\\
\tableline
\end{tabular}
\tablecomments{Calibration clusters used from Harris Catalogue (2010 edition) \citep{1996AJ....112.1487H} for globular clusters.  The Web version of the database for Galactic Open Clusters known as BDA (WEBDA) is used for the open cluster Melotte 66. $\mathrm{[Fe/H]_{lit}}$ gives the literature values for the cluster and $\mathrm{\Delta_{m}}$ is the difference between $\mathrm{[Fe/H]_{lit}}$ and $\mathrm{[m/H]_{mean}}$. The median is shown in the box plot of Figure \ref{fig:whiscal}.  The Num column gives the stars observed for each cluster.}
\end{center}
\end{table}
\footnotetext[6]{ The WEBDA metallicity listed is from \cite{1993A&A...267...75F} but all Harris catalog metallicities incorporate multiple literature values as found in the Harris bibliography tables.} 

The uncertainties for velocity measurements are known, from the AAOmega spectrograph used to take the observations,  to be 3 to 5 km s$^{-1}$ due to the spectral resolution of the 1700D AAOmega grating. To check the robustness of the velocities, and obtain a more precise error estimate, we compare the velocities of 21 different stars which were observed 3-4 times. We adopt the average error $\sigma/n\simeq 2.7$ km s$^{-1}$ as the error for the velocity measurements, $\mathrm{V_{err}}=2.7$ km s$^{-1}$. This average error will only be used for the purpose of drawing the generalized Gaussian histograms in  Section \ref{sec:mprob}.

\section{Sagittarius Stream Selection}
\label{sec:sgrsel}
 \begin{figure}[htb]
\begin{center}
\includegraphics[width=0.5\textwidth]{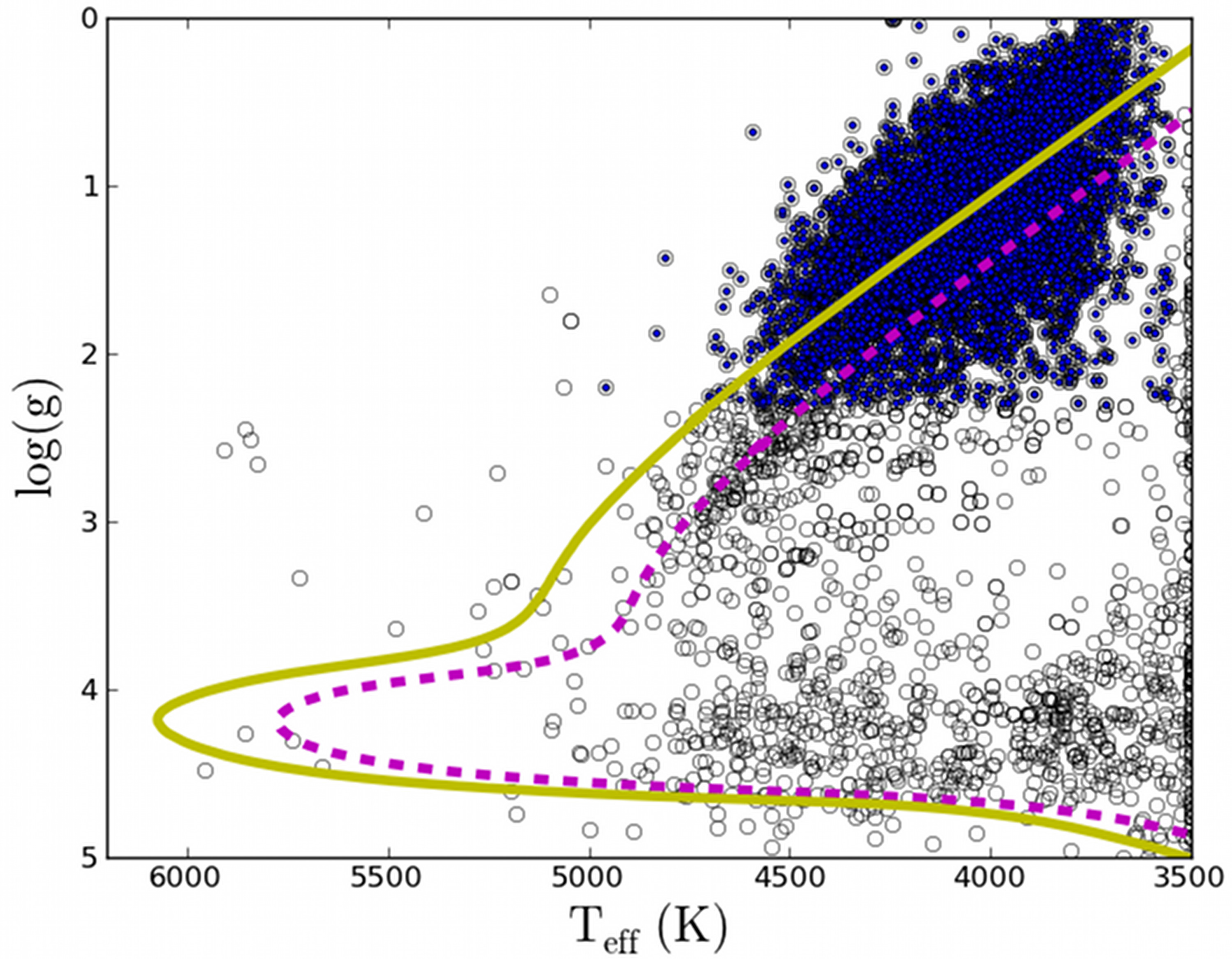}
\end{center}
\caption{The cuts in $\mathrm{\log(g)}$ and $\mathrm{T_{eff}}$ for F01. The filled circles represent the selected stars. The Dartmouth isochrones \citep{2008ApJS..178...89D} shown are a 10 Gyr isochrone at [Fe/H] $= 0.0$, as the dashed magenta line, and a 10 Gyr isochrone at [Fe/H] $=-0.5$, as the solid yellow line. Open circles are the stars before  selection.}
\label{fig:ccuts}
 \end{figure}

Using the modified RAVE pipeline we exclude targets with null or anomalous Ca II 8500 $\mathrm{\AA}$ triplet line measurements.  The RAVE   correlation coefficient $\mathrm{R_{coef}}$ determines the goodness of the fit between the star spectrum and the corresponding template through a standard cross correlation procedure \citep{1979AJ.....84.1511T}. Higher values of $\mathrm{R_{coef}}$ indicate a better fit, but the value of the cutoff depends on the effective temperature of the stars involved. For example, hotter stars will have intrinsically lower $\mathrm{R_{coef}}$ values.  For our data we use the quality selection on the RAVE correlation coefficient $\mathrm{R_{coef}}  \geq 15$,  and visually inspect the residual spectrum (the difference between the spectrum and the RAVE model spectrum). 

 As mentioned previously, the Sgr stream observations are divided into 29 regions (locations are given in Figure \ref{fig:sel1}). The K and M-giant stars are selected from RAVE $\mathrm{\log(g)}$ and $\mathrm{T_{eff}}$ values.
  
To separate only the K and M-giants from the data we select $\mathrm{\log(g)}$ and $\mathrm{T_{eff}}$ as follows: $\mathrm{\log(g)}$ $\leq$ 2.3  and $3500 \leq $$\mathrm{T_{eff}}$$ \leq 5000$. From previous studies of RAVE data  we adopt the standard RAVE cutoff, excluding metallicities of $\mathrm{[Fe/H] < -1.5}$ dex \citep{2011AJ....141..187S}, although our data calibration is consistent with a linear calibration down to [Fe/H] $\simeq -2.2$ in Figure \ref{fig:whiscal}.  

The selection of the F01 core region is shown in Figure \ref{fig:ccuts}. This high density region is the best illustration of our method and Figure \ref{fig:ccuts} shows the F01 Sgr selection compared to Dartmouth 10 Gyr isochrones for [Fe/H] $=0$ (dashed) and $-0.5$ (solid) metallicities \citep{2008ApJS..178...89D}.  The choice of 10 Gyr isochrones is motivated by previous fitting showing an age range of approximately 8-10 Gyrs in the Sgr dwarf stream \citep[e.g.,][]{2003ApJ...599.1082M,2006A&A...446L...1B,2010ApJ...714..229L,2010ApJ...712..516N,2012AJ....143...88C}. In F01 this selection returns 6,227 (filled circles) from the 9,086 candidate (open circles) stars.  Applying the same cuts in $\mathrm{T_{eff}}$ and $\mathrm{\log(g)}$ for the stream fields returns 809 candidates, giving a total of 7,036 candidate Sgr stars.  The pileup of stars seen near $\mathrm{T_{eff}} \sim$ 3500 K in Figure \ref{fig:ccuts} is an artifact of the modified RAVE pipeline, and they are excluded based on their low RAVE correlation coefficients.

For coordinates in this paper we adopt the Sgr coordinate system $\mathrm{\Lambda}$, B as defined by \cite{2003ApJ...599.1082M}. This system is oriented such that the equator is defined by the Sgr stream midplane, with the origin of $\mathrm{\Lambda}$ centered on the Sgr core. This flattens the stream along the sky, enhancing the visibility of features in the stream.

\subsection{Membership Likelihood}
\label{sec:mprob}
After excluding all but the  most likely candidates, it is still possible that the stars we have selected  are members of our own Milky Way halo rather than  being the desired Sgr stream stars. To identify Sgr stars in our sample we look for velocity substructure that deviates from the Galactic halo.

\begin{figure}
\begin{center}
\includegraphics[width=0.5\textwidth]{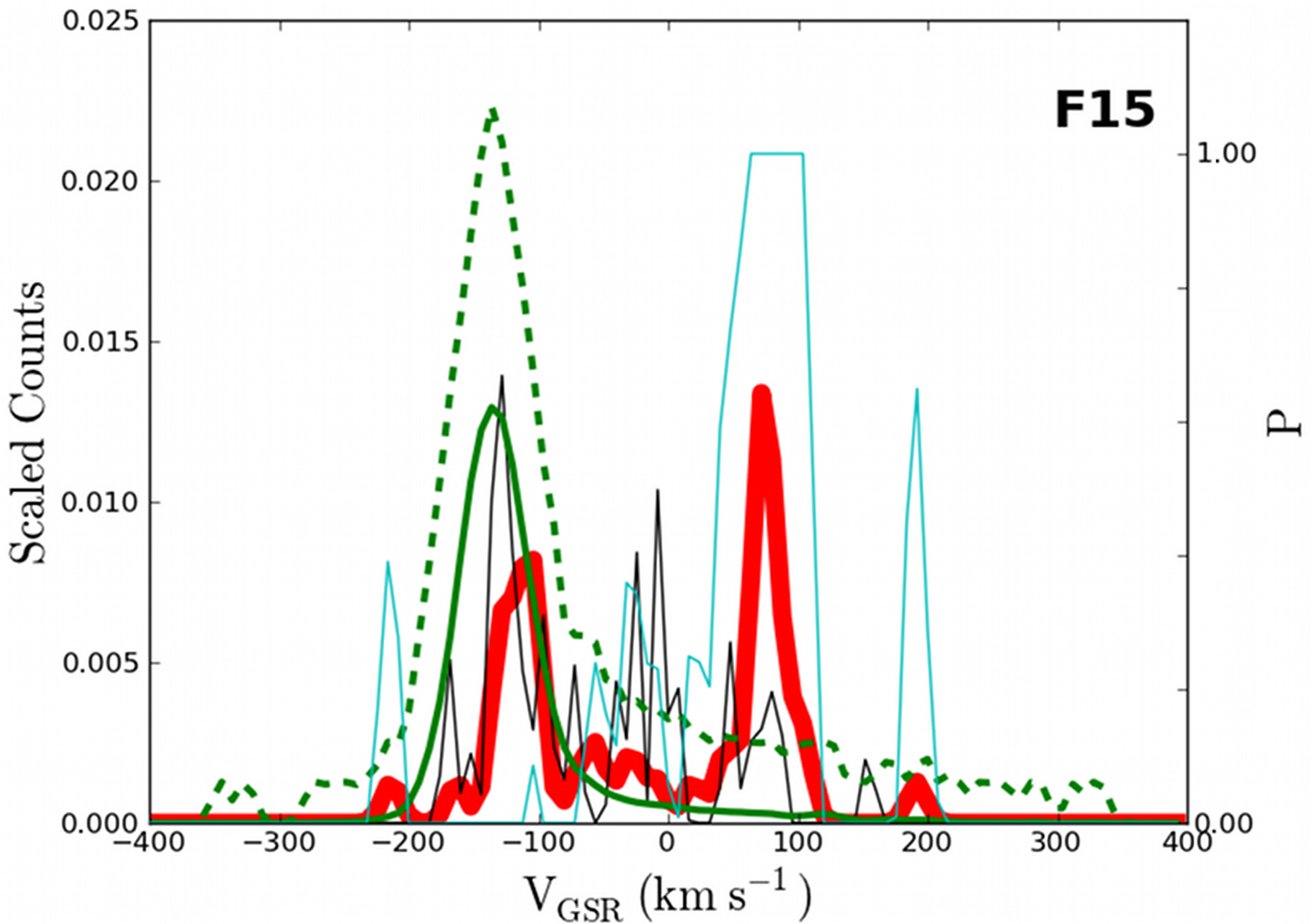}
\includegraphics[width=0.5\textwidth]{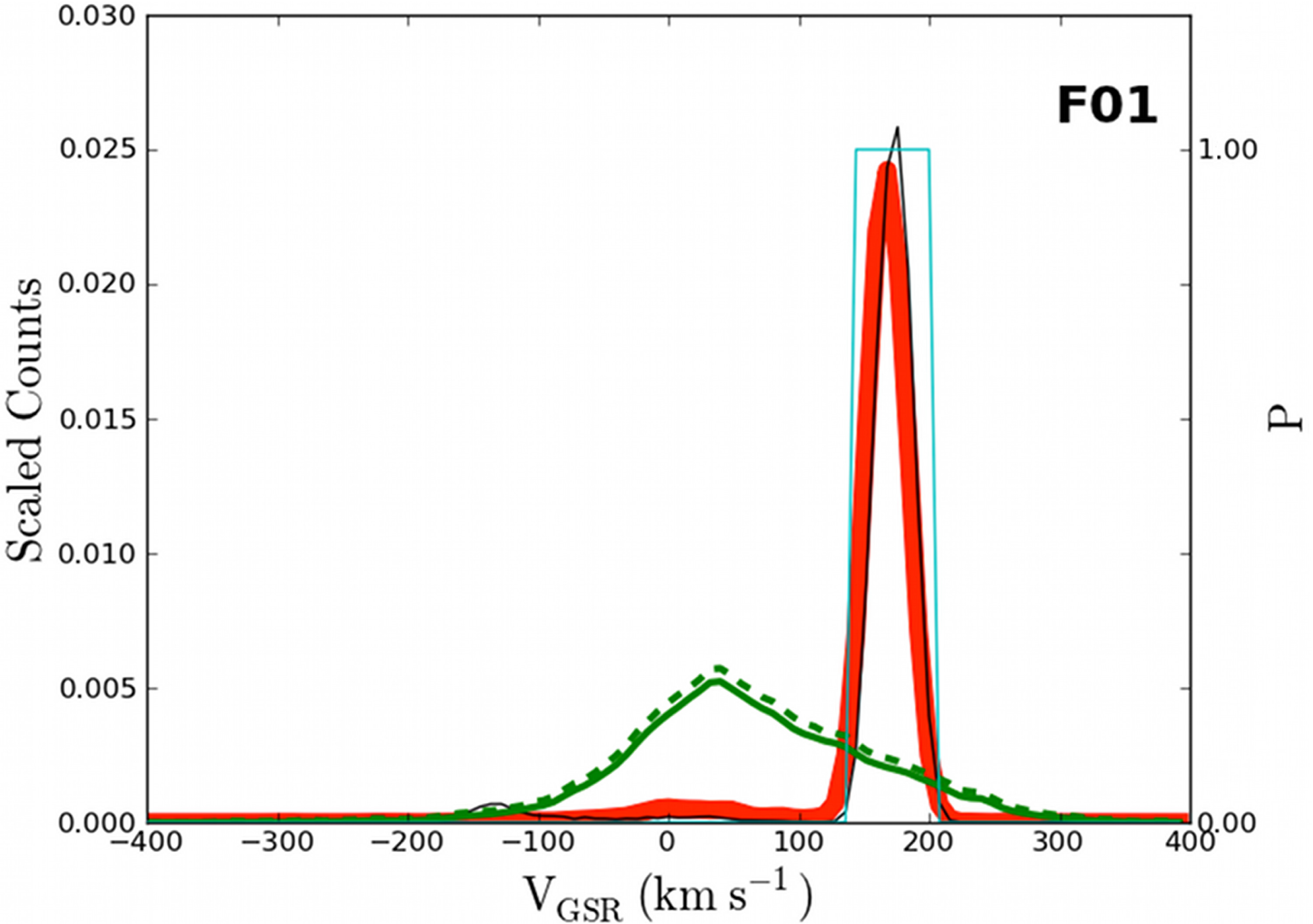}
\end{center}
\caption{The likelihood calculation for two regions, F01 and F15. The red thick solid line is the generalized Gaussian histogram for our data in the respective regions.  The scaled counts axis gives the amplitude of the  normalized distributions where we have set the distribution sampling to be identical to the data. The medium green solid line is the Besan\c{c}on model \citep{2003A&A...409..523R}.  The thin light blue colored line is the likelihood value P=$\mathrm{P_{NotHalo}}$ and the thin dark line is the \cite{2010ApJ...714..229L} model in the same region.  The top green dashed line marks the confidence level ($\mathrm{C_{bes}}$) cutoff for membership to the Besan\c{c}on distribution (the smooth Galactic halo).}
\label{fig:prob}
 \end{figure}

We use generalized Gaussian histograms to compare the data to the smooth Galactic halo and to define a selection criteria for likely Sgr members from the population of stars in a field. For each field, we consider the range of velocities between $\mathrm{V_{GSR}}=-400$ and $+$400 km s$^{-1}$.  Dividing that range into 100 bins we calculate the distribution:
\begin{equation}\label{eqn:dpop}
\mathrm{D_{field}=\sum_{i}^{N} D[i] = \frac{1}{\sqrt{2\pi}\sigma} e^{-(x-V)^{2}/(2\sigma^{2})}}
\end{equation}
over all bins, where i is each measurement $\mathrm{V_{GSR}}$ for each star in the field in question.  The value  V is given by each measurement of Galactocentric velocity ($\mathrm{V_{GSR}}$) in the field and N is the number of stars in the field. The value of $\sigma$ is the error associated with the velocity measurements.   The candidate distribution for each field is $\mathrm{D_{field}=D_{F01},.... D_{F30}}$.  We use the value of $\mathrm{V_{err}=2.7\ km\ s^{-1}}$ as defined in Section \ref{sec:analis} and  V $= \mathrm{V_{GSR}}$. This produces the red velocity distributions shown in Figure \ref{fig:prob}.

To compare our observations to the background of the smooth Galactic halo we use the velocity distribution as predicted by the Besan\c{c}on Galaxy model \citep{2003A&A...409..523R}.  Although there are other models available, to make our comparison we only need a general location of the halo in velocity space. We find the canonical Besan\c{c}on model is well suited to our purposes. We calculate the Besan\c{c}on generalized Gaussian histogram, $\mathrm{D_{B}}$, for each region we observe where the value of  V $= \mathrm{V_{vbc}}$, and $\mathrm{V_{vbc}}$ is $\mathrm{V_{GSR}}$ predicted by the model.  When using the Besan\c{c}on Galaxy model we select each region based on the center coordinates of our fields F01 to F30. \begin{figure*}[htb]
\begin{center}
\includegraphics[width=0.99\textwidth]{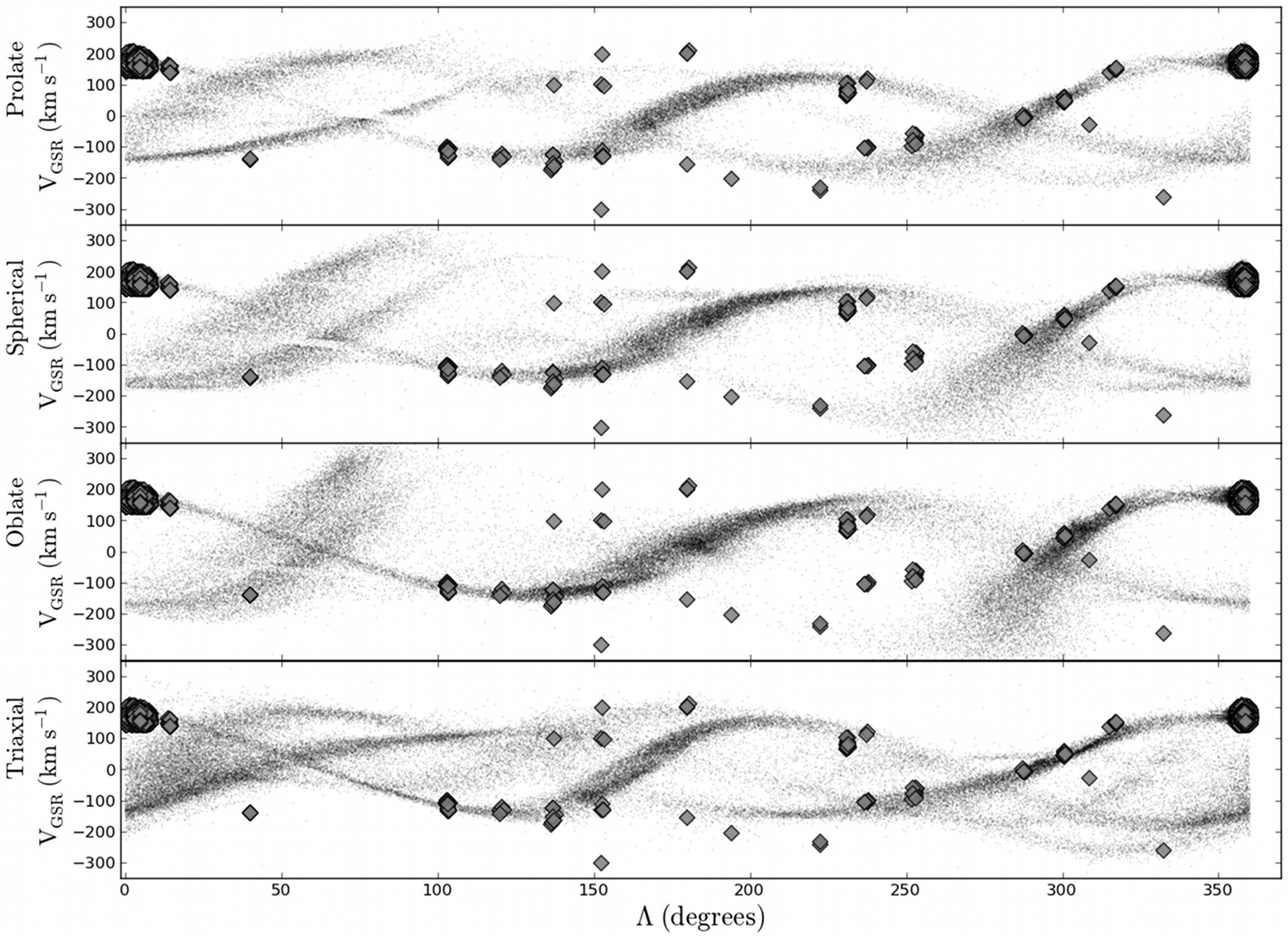}
\end{center}
\caption{ Comparison between our P1 data (diamonds) and the  prolate, spherical and oblate models from \cite{2005ApJ...619..807L}, as well as the triaxial model from \cite{2010ApJ...714..229L}. Both leading and trailing arm P1 debris kinematically overlap with the models. The prolate and triaxial models provides a better correspondence to the cluster of P1 points at $\mathrm{\Lambda\simeq250^{\circ}}$.}
\label{fig:modelcomp}
\end{figure*}We select stellar types to include M and K giants, and bright giants with the default parameters to represent the smooth Galactic halo. When we apply Equation \ref{eqn:dpop} to our Besan\c{c}on selection this results in $\mathrm{D_{B}}$, the green curve in Figure \ref{fig:prob}. This distribution is centered at the same location on the sky as a corresponding Sgr observation. For each $\mathrm{D_{field}}$ there is a corresponding $\mathrm{D_{B}}$ produced.

To calculate whether the star is potentially a member of Sgr we compare $\mathrm{D_{B}}$ with the  $\mathrm{D_{field}}$.  We also include in Figure \ref{fig:prob} the distribution from the triaxial model for the Sgr Stream \citep{2009ApJ...703L..67L,2010ApJ...714..229L} as the black line.  The black  distribution is drawn from the model points corresponding to the spatial region spanned (in degrees) by fields 1 - 30 respectively. We show F01 and F15 as examples of the output in Figure \ref{fig:prob}. The black line distribution is simply the generalized Gaussian histogram of the model points which lie within the field in question, i.e., for F01 this would be the model points within $\sim 6^{\circ}$ of the Sgr core.

The $\mathrm{D_{field}}$ and $\mathrm{D_{B}}$ distributions are used to calculate the likelihood of a given star in a region not being in the smooth halo. To establish limits on the likelihood we set stars with $\mathrm{D_{field} \leq D_{B}}$ to have a P$_{\mathrm{NotHalo}}=0$ and likewise if $\mathrm{D_{B}} \leq 0.001$ we let P$_{\mathrm{NotHalo}}=1$.  In the region where $\mathrm{D_{field} \geq D_{B}}$ then we need to consider the difference in the distributions.  To keep an accurate scale of likelihood between zero and one we will set a confidence level which can be used as a  cutoff in the region where $\mathrm{D_{field} \geq D_{B}}$.

For each field we have an appropriately large sample of the Besan\c{c}on model for that area. To determine the distribution for each Besan\c{c}on field (one corresponding to each of the 30 data fields)  we draw the Besan\c{c}on distribution 1,000 times using the same sample number of points as are present in the data (the red distribution).

Setting the confidence level at C=0.996 corresponds  to the level where 99.6\% of the data in the sample lies in 1,000 iterations of sampling the Besan\c{c}on model. The location of this confidence level gives us the cutoff $\mathrm{C_{bes}=C(0.996)}$. We choose this confidence level as it roughly corresponds to a $3\sigma$ cutoff for normally distributed data (but we do not assume anything about the distribution \emph{a priori}). The $\mathrm{C_{bes}}$ cutoff is drawn as the dotted green line in Figure \ref{fig:prob}.

All data in the red distribution which exceed $\mathrm{C_{bes}}$ (from the Besan\c{c}on model) are defined to be likely Sgr stars (rather than part of the smooth halo), so if $\mathrm{D_{field} \geq C_{bes}}$ then  P$_{\mathrm{NotHalo}}=1$.  If $\mathrm{D_{field} >  D_{B}}$ and $\mathrm{D_{field} \leq C_{bes}}$ we use:
\begin{equation}\label{eqn:pnoth}
 \mathrm{P_{NotHalo} = \left[ \frac{D_{field}}{D_{B}} - 1 \right]\frac{ D_{B}}{C_{bes}}}
\end{equation}
to calculate the likelihood.  Equation \ref{eqn:pnoth} statistically orders each data point between the limits of zero to one using the maximum and minimum defined by the confidence level $\mathrm{C_{bes}=C(0.996)}$. This gives us a smooth linear trend between the P$_{\mathrm{NotHalo}}=0$ (P0) and P$_{\mathrm{NotHalo}}=1$  limits.  Hereafter, all cases of $\mathrm{P_{NotHalo}=1}$ are collectively referred to as P1 events. The total P$_{\mathrm{NotHalo}}$ is shown (the light blue curve) with the scaled normalized data from the Besan\c{c}on model, $\mathrm{D_{B}}$, and the Sgr data, $\mathrm{D_{field}}$, as shown in Figure \ref{fig:prob}.  For each region $\mathrm{D_{field}}$ we then have some set of stars with velocity V that are included in the $\mathrm{P_{NotHalo}=1}$ subset and qualify as P1 events. Combining results from all of our regions we find a total of 5513 core F01 candidates and 106 stream candidates, a total of 5619 possible Sgr members.  Although we assume Sgr membership this may not be the case for all objects, as discussed further in Section \ref{sec:over}.

\section{Comparison of Models to the Sagittarius Stream}
\label{sec:compmod}

\footnotetext[7]{The data for these models are obtained from the website of David Law at: \texttt{http://www.astro.virginia.edu/ srm4n/Sgr/}}

Here we compare four models to the P1 selection of the Sgr stream. These four models\footnotemark[7] are generated with triaxial, spherical, prolate, and oblate Milky Way dark matter haloes, respectively. The spherical, prolate and oblate models were discussed in \cite{2005ApJ...619..807L} whereas the more recent triaxial model is from  \cite{2009ApJ...703L..67L} and \cite{2010ApJ...714..229L}.

Mapping the P1 stars returned by our selection technique, we find that the leading arm of Sgr \citep[known to be particularly sensitive both kinematically and spatially to the shape of the Milky Way dark matter halo;][]{2010ApJ...714..229L} overlaps with the predictions of both the triaxial and prolate models in velocity space, as shown in Figure \ref{fig:modelcomp}.  All four models show agreement in velocity as they pass through or near P1 points but only the prolate and triaxial models show agreement in the $\mathrm{\Lambda, V_{GSR} \simeq (250^{\circ}, -200\ km\ s^{-1})}$ region where we have a large number of high likelihood P1 Sgr stars.

 Though it is known that the triaxial model does not reproduce the bifurcation in the Sgr stream we consider it to be the best of these four options.  A potential solution  proposed by \cite{2012AJ....143...88C} is that this bifurcation may have resulted from a kinematic disruption of some kind, but if the bifurcation was caused by some more recent event then there would be no need to account for the bifurcation feature in formational models of the stream. However, there is still no model, disruptive or otherwise, that can account for all the observed Sgr stream properties.

\begin{figure*}[htb]
\begin{center}
\includegraphics[width=0.92\textwidth]{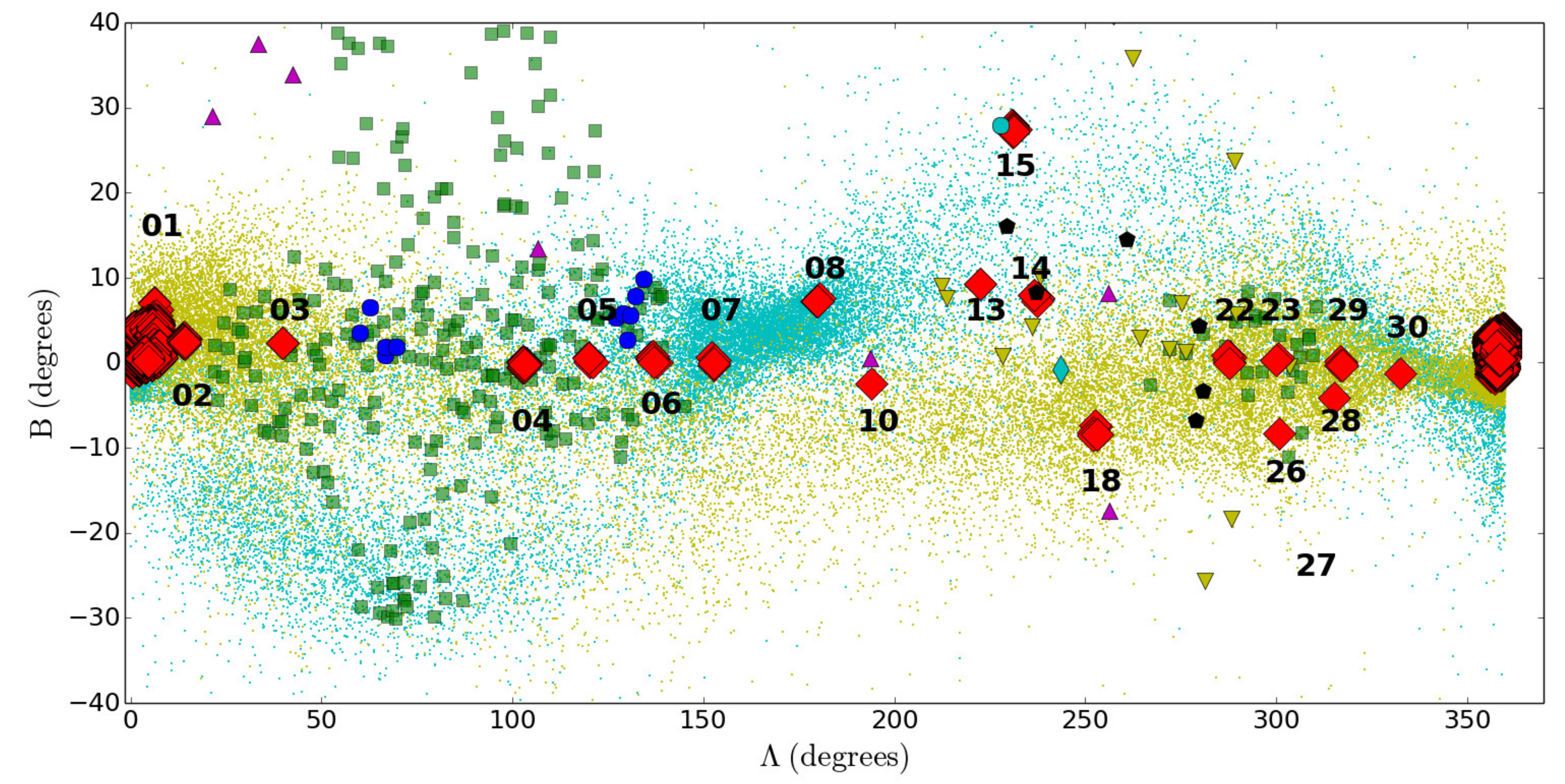}
\includegraphics[width=0.92\textwidth]{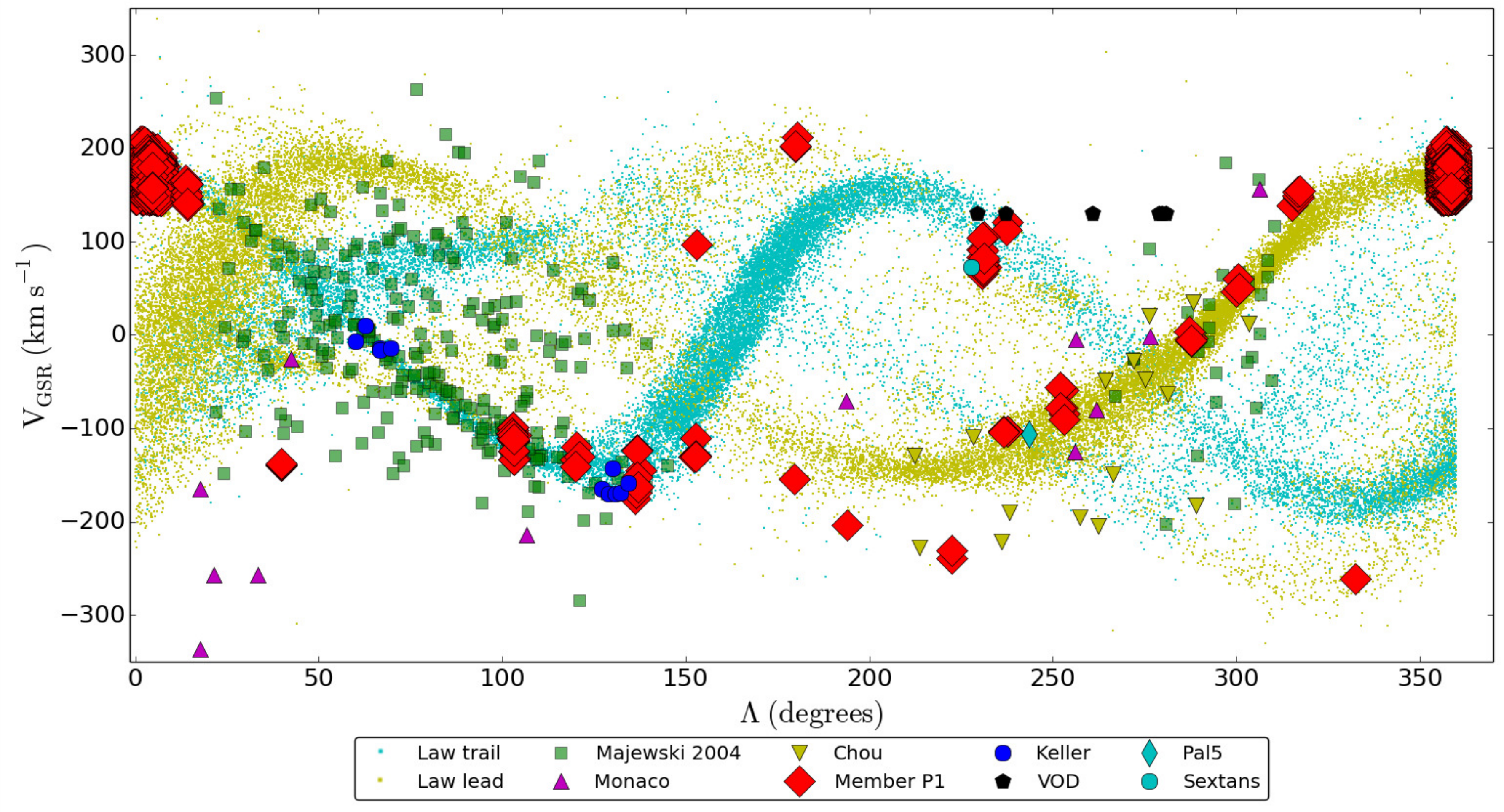}
\end{center}
\caption{The distribution of the highest likelihood Sgr stars. P1 stars are shown as large red diamonds, the Virgo Overdensity locations are shown as black pentagons,  Pal5 is shown as a thin blue diamond.  Top is the field numbering used in the likelihood calculations, starting at the core (F01) and going to stream field 30. For comparison we show the data from \cite{2004AJ....128..245M} as squares, the data from \cite{2010ApJ...720..940K} as circles, \cite{2007A&A...464..201M} as upwards triangles, and data from \cite{2007ApJ...670..346C} as downwards triangles. The triaxial model of \cite{2010ApJ...714..229L} is given by the very small dots  \citep[online data from models of][]{2009ApJ...703L..67L,2010ApJ...714..229L}. The trailing arm is shown in light blue and starts at a velocity of about 180 km s$^{-1}$ at $\Lambda=0$ and goes down to  $-150$ km s$^{-1}$ at 360 degrees, whereas the leading arm is shown in yellow and starts at $-150$ km s$^{-1}$ at $\Lambda=0$ and goes up to 180 km s$^{-1}$. }
\label{fig:sel1}
\end{figure*}

The triaxial model seems to match both leading and trailing arm observations kinematically, and as it is the most recent of the models we are considering, we use it as a point of comparison for our data in the stream. We show the full set of P1 coordinates and velocities as well as other observations from the literature with the predictions of the triaxial model in Figure \ref{fig:sel1}.

\section{Discussion}
\label{sec:dis}

 As shown in Figure \ref{fig:prob}, we have a range of calculated likelihood values that go from zero to one. The highest likelihood objects are P$_{\mathrm{NotHalo}}=1$ (P1) stars, which give us a total of 5,619 very high likelihood Sgr members, 106 of which are in the stream (i.e., not in F01). The P1 selection includes only those stars which lay on or outside of the $\mathrm{C_{bes}}$ dotted blue line shown in Figure \ref{fig:prob}.  
 
 We find that 100\% of the P1 stars selected in the F01 region lie within the known velocity range for the Sgr dwarf, i.e., within a range of heliocentric velocities between 100 and 200 km s$^{-1}$. This range is chosen based on the radial velocity profile of the core to within 3$\sigma$ \citep[from][$V_{core}=139.4 \pm 0.6$ km s$^{-1}$]{2008AJ....136.1147B}. Including all Sgr candidates for F01 we find 93.3\% of the stars are within the velocity range for Sgr.  The P1 selection then gives us zero halo objects in the well known Sgr core region, without assuming anything beyond that they are not part of the smooth Milky Way halo. We use these P1 stars to trace likely Sgr members in the stream where the velocity distribution is less well known. The spatial and velocity distribution of Sgr P1 data is compared to the triaxial model of \cite{2010ApJ...714..229L} in Figure \ref{fig:sel1}. We additionally show the positions and velocities observed for the stream from  \cite{2004AJ....128..245M}, \cite{2007ApJ...670..346C}, \cite{2007A&A...464..201M}, and \cite{2010ApJ...720..940K} for comparison.  
 
There are 106 P1 stream members shown in Figure \ref{fig:sel1}. The P1 data indicate that the distribution of Sgr stars in the core and stream approximately agree with the predictions from the triaxial model of  \cite{2010ApJ...714..229L}. In both leading and trailing arm regions of the model there is a general consistency in velocity space.  The discrepancy between data and models could be due to problems in the models or with the small number statistics, i.e., the low density of P1 stars in the stream. A list of the selected P1 stars is given in Table \ref{tab:p1}.

 \subsection{The Stream Metallicity Distribution}
 \label{sec:mdisstream}
We note agreement in velocity space for the P1 stream stars and the triaxial model  \citep{2010ApJ...714..229L}, listed as `Law' in the bottom panel of Figure \ref{fig:sel1}. Though there are a few interesting discrepancies (these may be due to other unknown overdensities or kinematic substructure, as discussed in Section \ref{sec:over}), the expectation is that the majority of P1 stars will be Sgr members. 

\begin{figure*}[ht!b]
\begin{center}
\includegraphics[width=0.92\textwidth]{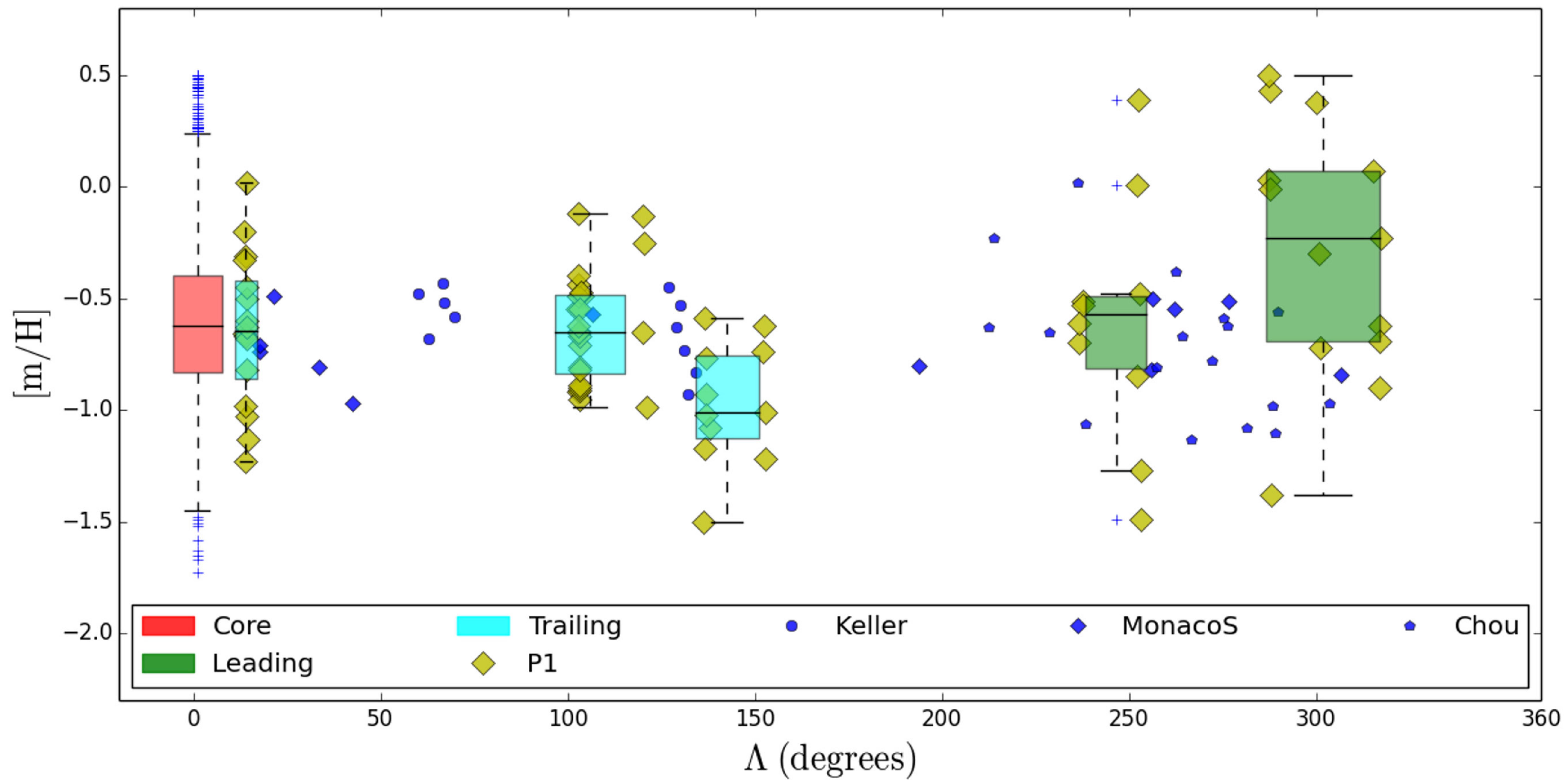}
\end{center}
\caption{  MDF evolution with $\Lambda$ for the leading (two negative $\Lambda$ boxes) and trailing (three positive $\Lambda$ boxes) Sgr arms.  The Sgr core (F01) region is shown by the single red boxed area at $\Lambda=0$. The black line in each box is the median value of metallicity for the area, the 25th percentile of the data is given by the upper edge of the box, the 75th percentile is given by the lower edge of the box and the maximum and minimum are at the ends of the whiskers.  Our Sgr points included are shown as yellow diamonds, comparison data are shown from \cite{2010ApJ...720..940K} as small blue circles, \cite{2007A&A...464..201M} as diamonds, and \cite{2007ApJ...670..346C} as pentagons.}
\label{fig:mdist}
\end{figure*}

The largest discrepancies between the data and the triaxial model lie in the region $\Lambda \sim 150^{\circ} - 250^{\circ}$ shown in Figure \ref{fig:modelcomp} and Figure \ref{fig:sel1}.  The triaxial model trailing arm starts at a velocity of about 180 km s$^{-1}$ at $\Lambda=0$ and goes down to a velocity of  about  $-150$ km s$^{-1}$ at 360 degrees whereas the leading arm starts at $-150$ km s$^{-1}$ at $\Lambda=0$ and goes up to 180 km s$^{-1}$. The areas with the highest density of P1 stars show a good correspondence with the model but the mis-matches indicate that there is more work to be done in particular in fitting the trailing arm. Towards that end we investigate the metallicity distribution from the core and across the stream.

We find the average metallicity of core (F01) P1 stars to be $-0.59$ dex with a dispersion of $\sigma \simeq 0.34$ dex.   This is only slightly more metal poor than the previously measured mean of [Fe/H]$\simeq-0.5$ dex \citep{2005A&A...441..141M,2010ApJ...720..940K,2002IAUS..207..168C},  and we expect that the metal poor tail will extend quite a bit lower and indeed may even go into the extremely metal poor range, i.e. having metallicities of $\mathrm{[Fe/H] < -3.0}$.

Without selecting purely for  M-giants we have about 30 stars with $[$m/H$]\leq-2.0$ dex in our pre-processed F01 data. These stars have velocities and positions consistent with the central region of the Sgr dwarf (within $\sim 6^{\circ}$ of the core), and these intriguing objects are part of an ongoing follow-up campaign \citep{2012ASPC..458..325H}.  The RAVE errors on many of these measurements do not pass the selection criteria for this paper but the spectra are sorted via an individual \emph{splot} IRAF measurement of the Ca II 8500 $\mathrm{\AA}$ triplet spectral lines.

If our 30 potentially metal poor Sgr objects can be confirmed with high resolution spectroscopy, this number of metal poor objects in F01 would support the findings of \cite{2012AJ....143...88C}, who indicate that the distribution in the core of Sgr may be more metal poor than was previously thought.  As stated previously, for this publication we disregard any poor fits to the RAVE templates.  This means many of the very lowest metallicity stars fall out of the reduction, as they are not measurable by the automated routines we used.


To investigate the leading and trailing arm of the stream we select stream stars that are coincident with the triaxial model in the $\Lambda$, B, and $\mathrm{V_{GSR}}$ coordinates.  We also require that the stars not be in an overlapping region, i.e., areas of the model where both leading and trailing stars are predicted to lie. In Figure \ref{fig:mdist} we show the bins containing only P1 points. \cite{2005ApJ...619..807L}  note that along the leading arm of the stream, stars which are lost from Sgr in different orbits around the Milky Way can overlap in orbital phase position.  Possible effects of angular phase smearing and populations of stars from different evolutionary phases is given in \cite{2010ApJ...714..229L}. As mentioned by \cite{2007ApJ...670..346C} and \cite{2010ApJ...720..940K}, the longer trailing arm yields better energy sorting of the debris and can be more cleanly isolated from background or mixed populations. To select a clean set for leading and trailing arm populations we choose only stars which do not occupy a mixed phase position and create the leading and trailing population as shown in Figure \ref{fig:mdist}.

 In Figure \ref{fig:mdist} we create 2 bins for the leading arm (green) and 3 bins for the trailing arm  (light blue) along 360 degrees of $\Lambda$. These bins are defined to maximize the number of stars included as well as the spatial resolution as shown in Figure \ref{fig:mdist}. The five stream bins are located at $\mathrm{\Lambda \simeq [301,246,142,106,14]}$ with a range of $\mathrm{\Lambda_{min} - \Lambda_{max}} \simeq [30,17,17,18,1]$ and $[13,10,11,24,16]$ stars in each bin respectively. The distributions in metallicity in these bins are shown in Figure \ref{fig:mdist}, with the median of the bin given by a horizontal line. The mean metallicity of the five stream bins are $[-0.27,-0.60,-0.97,-0.63,-0.64]$ dex, with MDF dispersions of $[0.55,0.52,0.26,0.24,0.33]$ dex corresponding to the $\Lambda$ coordinates above.  The core F01 region has a mean metallicity of $-0.59$ dex and a dispersion of 0.34 dex from our sample of 5513 P1 core stars. The P1 objects which do not exclusively lie within a leading or trailing arm portion of the model are not included in the bins.

We find that the mean metallicity starting from $-0.59$ dex in the core, drops from $-0.63$ dex to $-0.97$ dex as $\Lambda$ increases along the trailing arm in Figure \ref{fig:mdist} (in the positive direction).  In the leading  arm, however, there is little or no trend with $\Lambda$ for the metallicity. The higher variation in metallicity values in the leading arm may be preventing any discernment of trends. Mapping of the arms by \cite{2007ApJ...670..346C} indicates that leading arm stars seem to be more metal poor than the core. In our data as well as that of \cite{2007ApJ...670..346C} the difference in the mean metallicities of the samples is less than their MDF dispersions (0.31 and 0.33 dex respectively), indicating that more detections will be needed before a gradient can be considered to be well established.

Kinematics of red horizontal branch stars from \cite{2012ApJ...751..130S} suggest a metallicity gradient of $\mathrm{-(1.8 \pm 0.3)\times10^{-3}\ dex\ deg^{-1}}$ in the trailing arm and a smaller gradient of $\mathrm{-(1.5 \pm 0.4)\times10^{-3}\ dex\ deg^{-1}}$ in the leading arm. Although these methods are different from the ones used in this paper, such a gradient does agree with theories for the formation of the dwarf galaxy where a gradient is initially present. In  particular, the more metal poor stars shown in our trailing arm data agree with the scenario of \cite{2012ApJ...751..130S} wherein the Sgr dwarf galaxy would have been formed with a more metal rich core and would have been surrounded by older and more metal poor stars. Those more metal poor stars would then have been stripped first and found further along the stream, creating the gradient we see.

\citet{2007ApJ...670..346C} find that the MDF of the Sgr stream changes from a median of [Fe/H]$\simeq-0.4$ dex to about $-1.1$ dex over a leading arm length, but our values do not drop off at that level, staying in the range of [m/H]$\simeq-0.61$ dex to about $-0.27$ dex.   Though the metallicity distribution in leading arm stars is known to be quite broad we find that our leading arm stars seem to lack a decrease to metal poor, and do not seem to show more metal poor stars than the trailing arm as mentioned by \cite{2007ApJ...670..346C}.  The broad range of metallicities in the leading arm has additionally been noted by \cite{2012AJ....143...88C}. The median metallicity of the Galactic thick disk is $\sim -0.7$ dex,  and, as noted by \cite{2007ApJ...670..346C}, should not have a large impact on MDF trends across the stream.  The trend to lower metallicities in the leading arm of the stream found by \cite{2007ApJ...670..346C}  may be related to their slightly more metal rich value for the Sgr core. 

\begin{table*}[h!tb]
\begin{center}
\caption{The approximate position and velocity of P1 overdensities.\label{tab:ovd}}
{\scriptsize
\begin{tabular}{cccccccccccccc}
\tableline
 Field &Stars& $\mathrm{\Lambda_{avg}}$&$\mathrm{B_{avg}}$&$\mathrm{RA_{avg}}$&$\mathrm{DEC_{avg}}$&[m/H]$_{avg}$&$\mathrm{\sigma_{mH}}$&$\mathrm{V_{GSR}}$(avg)&$\mathrm{\sigma_{v}}$ & catagory\\
\tableline
 &&degrees&&degrees& & dex &dex&(km s$^{-1}$)&(km s$^{-1}$)&\\
\tableline
01	&5513	&  154.32		&1.65  		&19.0	&-30.8	 	&-0.59	&0.34 	&168.76	&12.15 & Core F01 region of Sgr\\
04	&23		&  103.05		&-0.22  		&02.0	&-01.9	     &-0.65	& 0.2		&-112.61	&10.90  &Potential Sgr Stream clump\\
15	&17		&  231.02		&27.52  	&10.2	&-01.4	   	&-1.10,+0.48		&0.65 	&79.45	&11.49  &  Potential Sextans detection\\
23	&3		&  300.31		&0.23 		&15.0	&-06.9 	 	& 0.38	& 0.02	&50.71	&5.39   & Potential Sgr Stream clump\\
\tableline
\end{tabular}
}
\tablecomments{These groups correspond only to collections of three or more stars within $\pm$25 km s$^{-1}$ in velocity for an area with a 5 degree radius on the sky.  The distribution of all points is given in Figure \ref{fig:sel1}.}
\end{center}
\end{table*}

The comparison of the Sgr core region with the leading arm does not show the decline in median metallicity which would be represented by debris lost some 3.5 orbits ($\sim2.5-3$ Gyr) ago \citep{2005ApJ...619..807L}, but we do find substantial variation in the MDF along the stream. The observed MDF variation supports the theory  of \cite{2004ApJ...601..242M}, who suggest the satellite shed successive layers in its orbit, over which there must have been an intrinsic MDF gradient.  The disagreement in the direction of the trend between our data and that of \cite{2007ApJ...670..346C} indicates that perhaps the variation may be wider, and the gradient less, than either of our samples suggest.  This scenario would support models in which a rapid change in the binding energy of Sgr occurred over the past several gigayears, providing a large net metallicity variation, but a shallow gradient. The sudden change of state could have been caused by some dramatic event, which may also provide the source of kinematic disruption invoked by \cite{2012AJ....143...88C} to account for the bifurcation seen in the stream.

We additionally find a large MDF variation in the trailing arm sample, as mentioned above.  If this region is truly less susceptible to overlap in orbital phase position  then this would indicate evolution in the stream from a mean of -0.63 (near what we find for the core) to -0.97 dex (our most distant trailing arm box). This variation is additionally on the order of the MDF dispersion, but with overall lower dispersion than what is found in the leading arm.  We then can conclude that although the dispersion in the leading arm may be due to mixed orbital phase, the trends we find in the leading and trailing arms support a change in the Sgr MDF with position along the stream, and therefore an intrinsic MDF gradient in the progenitor to the Sgr dwarf and stream system we see today.

\subsection{Overdensities in the Stream}
\label{sec:over}


In Figure \ref{fig:sel1} we include information for known features (overdensities and streams) that are near the Sgr stream at various points on the sky. While we don't target them in our observations, there are several which have noteworthy overlaps.

Firstly we consider the Virgo Overdensities (VOD). Discovered independently by \cite{2001ApJ...554L..33V,2002ApJ...569..245N,2010PASA...27...45K} and \cite{2012EPJWC..1902007V}, the VOD includes three halo substructures: the two structures at an Right Ascension of 160 and 180 degrees respectively (at distances of 17 and 19 kpc, and with radii of 1.3 and 1.5 kpc) and an extended feature at 28 kpc that covers at least 162 deg$^{2}$ (the Virgo Equatorial Stream).  Perhaps covering 1000 sq deg or more on the sky \citep{2008ApJ...673..864J, 2012AJ....143..105B}.  We use the coordinates found in \cite{2010PASA...27...45K} to plot the spatial points and we adopt $\mathrm{V_{GSR}}=130\pm10$ km s$^{-1}$ for the VOD from \cite{2007ApJ...668..221N}.  This gives us  the black pentagons shown in Figure \ref{fig:sel1}.

The Palomar 5 (Pal5) globular cluster is located at coordinates l, b $\sim (0.85^{\circ}, 45.86^{\circ})$ shown by the light blue diamond. The metallicity is [Fe/H]=$-1.41$ dex from the Harris Catalogue \citep[2010 edition;][]{1996AJ....112.1487H} which is only slightly richer than the Sgr core region. Pal5 has an extremely low velocity dispersion, with a heliocentric velocity $v_{hel}=-58.7 \pm 0.2$ km s$^{-1}$ and a total line-of-sight velocity dispersion of $1.1 \pm 0.2$ km s$^{-1}$ \citep{2002AJ....124.1497O}.  From Equation \ref{eqn:v} this translates into $\mathrm{V_{GSR}}\sim -106.47$ km s$^{-1}$.

Finally we consider the Sextans dwarf, centered on the light blue circle. At coordinates of (RA, Dec) $\simeq (10h13',\ -01^{\circ}36')$  or (l,b) $ \simeq (243^{\circ}.5,\ 42^{\circ}.3)$ the Sextans dwarf is known to have a Galactocentric velocity of $\mathrm{V_{GSR}}\simeq$73 km s$^{-1}$  \citep{1990MNRAS.244P..16I,1995MNRAS.277.1354I,2004AJ....127.2031K}.  Sextans members have been measured to have a metallicity of [Fe/H]$=-1.7 \pm 0.25$ dex \citep{1991MNRAS.249..473D} which is slightly more metal poor than what we expect for Sgr stars. With a diameter of 30 arcminutes on the sky, this large and somewhat diffuse structure is similar to the type of detection we are looking for in the Sgr stream.

 The structure of Sgr overlaps with these overdensities in several regions  and it appears to remain well separated when metallicity, velocity, and spatial coordinates are taken into account, except in the case of the Sextans dwarf. The presence of the comparison features above led us to search for additional groupings of stars in our data. 

\begin{figure}[htb]
\begin{center}
\includegraphics[width=0.5\textwidth]{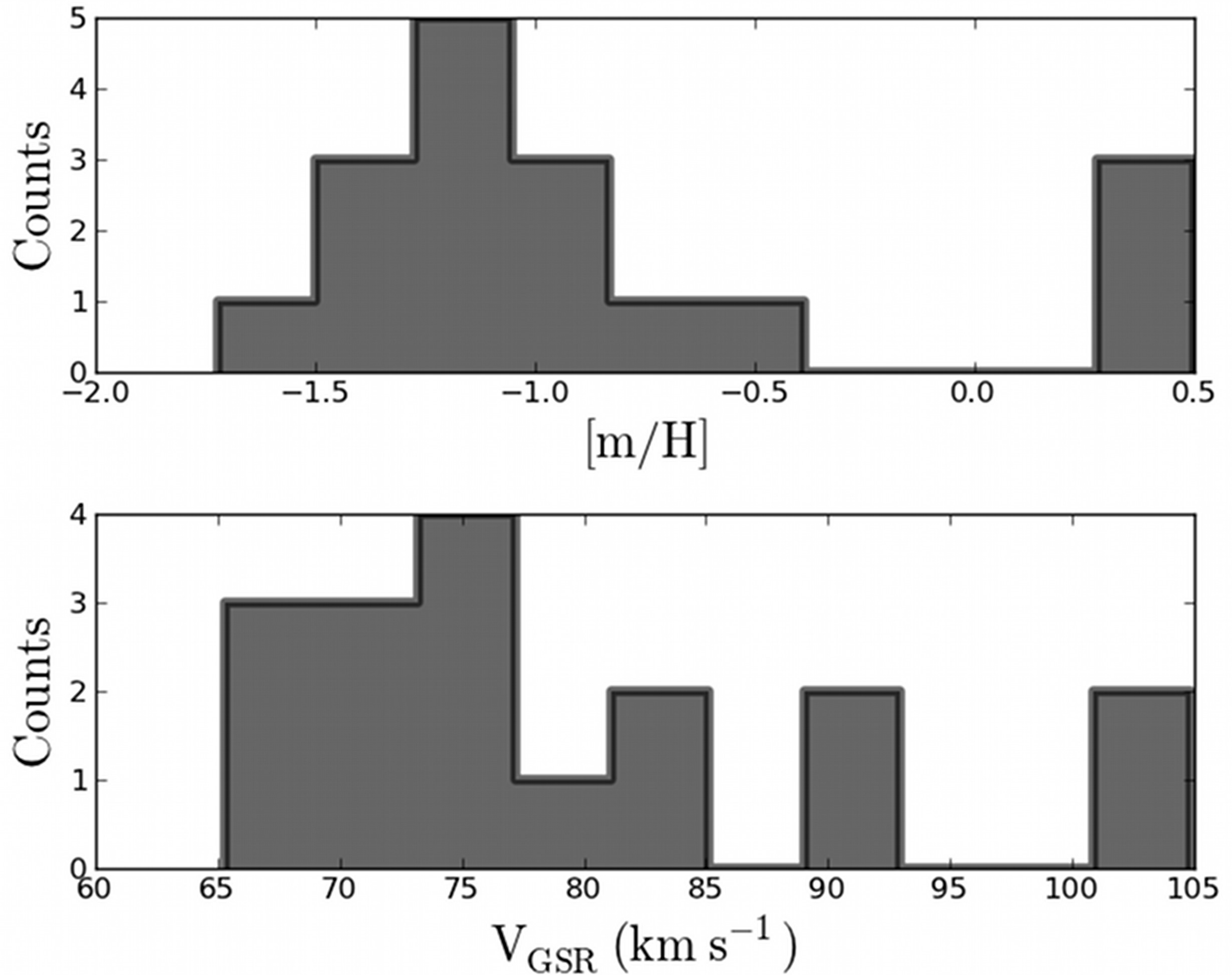}
\end{center}
\caption{The top panel plots the MDF for the F15 overdensity, showing a super-solar population as well as a lower metallicity group at [m/H]$\simeq -1.10$ dex.  The bottom panel shows the velocity distribution for the F15 overdensity; the higher velocity objects do not correspond to the super solar metallicity group. The detection of P1 stars for F15 in Figure \ref{fig:prob} translates to the above ranges in [m/H] and $\mathrm{V_{GSR}}$.}
\label{fig:vdfcomp}
\end{figure}

We searched the P1 data and found 4 regions which have stars that are coincident in velocity and spatial coordinates; i.e. within $\pm$25 km s$^{-1}$ in velocity for an area with a 5 degree radius on the sky. These sets or `over dense' regions in the data are located in F01, F04, F15 and F23 as shown in Table \ref{tab:ovd}.  While F01,F04, and F23 velocities correspond with what is expected from predictions of the triaxial model, the group of $\sim 17$ stars from F15 has a large offset with respect to the triaxial model. This detection is easily observed in Figure \ref{fig:prob}, where comparison shows little correspondence with the Milky Way smooth halo or the triaxial model for Sagittarius.  The peaks for the MDF and the velocity distribution of F15 are given in Figure \ref{fig:vdfcomp}. 

The group of 17 stars in F15 has an average velocity of $\mathrm{V_{GSR} = 79.45\ km\ s^{-1}}$ (where $\sigma=11.49$ km s$^{-1}$) and an average metallicity of [m/H]$_{avg} = -0.83$ (where $\sigma= 0.65$ dex) for a position centered on $\Lambda$, B $\simeq(231.02^{\circ}, 27.52^{\circ})$.  From Figure \ref{fig:vdfcomp}  the MDF seems to be bimodal; however, the three supersolar metallicity values do not reflect a grouping in velocity.  The coordinates in ([m/H],$\mathrm{V_{GSR}}$) are (+0.39,83.79), (+0.5,65.31), and (+0.5,103.94) and removing them from our distribution will shift the mean metallicity to [m/H]$\simeq -1.10$ dex but will have almost no effect on the average velocity.

The F15 group is near the spatial coordinates of three of our overdensities, namely the  VOD and the Sextans dwarf. However, the velocity measured corresponds closely only with Sextans.  As mentioned previously, we expect a value of approximately 73 km s$^{-1}$ from the literature for Sextans, and we measure $\mathrm{V_{GSR} = 79.45\ km\ s^{-1}}$ for F15.  This is dissimilar to the other overdensities in the region, i.e., the adopted VOD velocity ($\mathrm{V_{GSR}} \simeq$ 130 km s$^{-1}$)  and the expected velocity of the Sgr stream from the triaxial model (which shows peaks at $\mathrm{V_{GSR}} \simeq$ 0 and $-150$ km s$^{-1}$ as shown in Figure \ref{fig:prob}). This potential detection of Sextans in our data provides an additional validation for our statistical method and its ability to distinguish stellar structures from the smooth Galactic halo. The diffuse nature of Sextans provides us with an additional object which has an extended nature and suffers from foreground star contamination. That we can detect this object with a strong signal lends confidence to our other detections as truly being `\emph{not}' part of the Milky Way halo population.

\section{Conclusions}
\label{sec:conc}
We introduce a selection technique to separate Sgr member stars from the background.  As part of this method we first select only the best K and M-giant measurements from 24,110 spectroscopic observations. We then assign a likelihood of a given star \emph{not} being part of the smooth Galactic halo.  This is done by drawing a distribution using Equation \ref{eqn:dpop} and calculating a likelihood per star per field using Equation \ref{eqn:pnoth}. We identify 106 likely members of the Sgr stream and find three main results:
\begin{enumerate}
\item{The Sgr stream $\mathrm{V_{GSR} }$ distribution with $\Lambda$ shows reasonable agreement with a triaxial model for the Milky Ways dark matter halo. We compare the kinematics of the observed Sgr stream stars with those of extant simulations of the tidal disruption of Sgr.  From comparisons with 4 models we identify the triaxial model of \cite{2010ApJ...714..229L} as a reasonable fit to the data and find general agreement across the Sgr stream.}
\item{The Sgr stream MDF variation with $\Lambda$ has yielded several interesting results. We find that the trailing arm mean metallicity seems to become more negative further away from the Sgr core.  We measure a decrease in mean metallicity from $-0.59$ dex in the core to $-0.97$ dex with increasing $\Lambda$. This metallicity gradient, in which material further from the Sagittarius core is more metal-poor, is consistent with the scenario of tidal disruption from a progenitor dwarf galaxy that possessed an internal metallicity gradient.  In the leading arm, however, we have larger variations in metallicity values, lacking a clear trend; this larger range of values may be due to a mixing of orbital phases in the leading arm.}
\item{A search for overdensities finds three groups consistent with the triaxial model  (F01, F04, F23) and one potential detection of the Sextans dwarf at F15. We report on this new detection, the F15 overdensity, located at RA, DEC $\simeq$ (10.2, $-$01.4), which has peaks in the metallicity distribution at [Fe/H] $= -1.10$ and $+0.46$ dex and an average velocity of  $\mathrm{V_{GSR}=78.48\ km\ s^{-1}}$. The F15 overdensity does \emph{not} appear to coincide with either the triaxial model of \cite{2010ApJ...714..229L}, the  VOD or the smooth Galactic halo as represented by the Besan\c{c}on model; but it does coincide in both spatial and velocity coordinates with the Sextans dwarf. }
\end{enumerate}
The parameter space we use for this study includes the modified RAVE pipeline $\mathrm{\log(g)}$, $\mathrm{T_{eff}}$, [m/H] metallicity, and velocity. We obtained velocity information through fitting the Ca II 8500 $\mathrm{\AA}$ triplet with \emph{IRAF} software and we use the 2MASS photometric colors for initial object selection,  but for many of our objects this parameter space could be expanded. The \emph{ppmxl} \citep{2010AJ....139.2440R} catalog, the Sloan Digital Sky Survey (SDSS) \citep{2000AJ....120.1579Y} observations, and the ongoing Skymapper photometric survey \citep{2012ASPC..458..413C} may all be incorporated in the future.

{\scriptsize
\begin{longtable*}{ccccccccccc}
\caption{The selected 106 likely Sagittarius stream members with modified RAVE pipeline metallicity, surface gravity and effective temperature values.\label{tab:p1}}\\
\hline
Num & $\mathrm{RA_{J2000}}$ & $\mathrm{DEC_{J2000}}$ & $\mathrm{\Lambda}$ &  B & log(g) & $\mathrm{T_{eff}}$ & $\mathrm{[m/H]}$ &  Rcoeff \\
 & &  & (degrees) & (degrees) & &(K) & (dex)  &  \\
\hline
\hline
\endfirsthead
\multicolumn{10}{c}%
{\tablename\ \thetable\ -- \textit{Continued from previous page}} \\
\hline
Num &$\mathrm{RA_{J2000}}$  & $\mathrm{DEC_{J2000}}$ & $\mathrm{\Lambda}$ & B & log(g) & $\mathrm{T_{eff}}$ & $\mathrm{[m/H]}$  &  Rcoeff \\
\hline
\endhead
\hline \multicolumn{4}{r}{\textit{Continued on next page}} \\
\endfoot
\hline
\endlastfoot
\hline
  1      &  01:58:05.38 & -01:45:11.2 &  102.62 &  -0.58 &  1.31    &  4249.0  &  -0.92 &  21.21     \\
  2      &  01:58:37.34 & -01:44:19.0 &  102.75 &  -0.52 &  1.18    &  4182.0  &  -0.49 &  28.49     \\
  3      &  01:58:37.34 & -01:44:19.0 &  102.75 &  -0.52 &  1.09    &  4209.0  &  -0.65 &  31.0      \\
  4      &  01:58:45.53 & -01:43:24.2 &  102.78 &  -0.52 &  1.48    &  4220.0  &  -0.4  &  23.83     \\
  5      &  01:58:45.53 & -01:43:24.2 &  102.78 &  -0.52 &  1.34    &  4148.0  &  -0.62 &  22.76     \\
  6      &  01:59:10.97 & -02:19:21.7 &  102.57 &  0.05  &  1.66    &  4571.0  &  -0.55 &  36.68     \\
  7      &  01:59:50.95 & -02:08:49.6 &  102.8  &  -0.02 &  1.08    &  4098.0  &  -0.12 &  20.4      \\
  8      &  02:00:07.30 & -02:23:16.8 &  102.74 &  0.23  &  1.36    &  4206.0  &  -0.44 &  31.64     \\
  9      &  02:00:10.75 & -01:39:51.5 &  103.12 &  -0.39 &  1.29    &  4470.0  &  -0.71 &  30.25     \\
  10     &  02:00:10.75 & -01:39:51.5 &  103.12 &  -0.39 &  1.66    &  4511.0  &  -0.9  &  36.04     \\
  11     &  02:00:33.24 & -01:36:46.4 &  103.23 &  -0.39 &  1.75    &  4368.0  &  -0.81 &  21.76     \\
  12     &  02:00:33.24 & -01:36:46.4 &  103.23 &  -0.39 &  1.95    &  4330.0  &  -0.89 &  23.38     \\
  13     &  02:00:36.79 & -02:05:40.9 &  102.99 &  0.03  &  1.16    &  4378.0  &  -0.65 &  39.38     \\
  14     &  02:00:36.79 & -02:05:40.9 &  102.99 &  0.03  &  0.91    &  4352.0  &  -0.91 &  29.51     \\
  15     &  02:01:00.19 & -01:48:00.0 &  103.23 &  -0.17 &  2.2     &  4674.0  &  -0.67 &  24.61     \\
  16     &  02:01:00.19 & -01:48:00.0 &  103.23 &  -0.17 &  1.94    &  4891.0  &  -0.82 &  25.6      \\
  17     &  02:01:03.12 & -02:02:04.2 &  103.12 &  0.04  &  0.27    &  3926.0  &  -0.95 &  41.44     \\
  18     &  02:01:03.12 & -02:02:04.2 &  103.12 &  0.04  &  1.67    &  4304.0  &  -0.55 &  34.22     \\
  19     &  02:01:16.68 & -01:34:20.6 &  103.4  &  -0.33 &  1.78    &  4279.0  &  -0.47 &  21.67     \\
  20     &  02:02:33.29 & -01:38:29.4 &  103.64 &  -0.11 &  2.22    &  4505.0  &  -0.49 &  29.09     \\
  21     &  03:00:12.67 & +05:50:44.5 &  119.85 &  0.69  &  1.62    &  3836.0  &  -0.65 &  39.23     \\
  22     &  03:00:20.59 & +06:02:06.7 &  119.97 &  0.54  &  2.15    &  4404.0  &  -0.13 &  15.83     \\
  23     &  03:00:40.32 & +06:34:28.6 &  120.31 &  0.11  &  1.84    &  4679.0  &  -0.25 &  27.24     \\
  24     &  03:02:50.71 & +06:53:02.0 &  120.93 &  0.11  &  0.82    &  4116.0  &  -0.99 &  34.77     \\
  25     &  03:58:20.57 & +13:42:56.9 &  136.18 &  0.64  &  1.11    &  4559.0  &  -1.5  &  39.35     \\
  26     &  03:58:45.77 & +13:58:35.4 &  136.39 &  0.46  &  0.54    &  3791.0  &  -1.17 &  43.81     \\
  27     &  03:59:42.72 & +14:07:32.5 &  136.66 &  0.43  &  1.33    &  4020.0  &  -0.59 &  19.55     \\
  28     &  04:00:24.00 & +14:40:31.1 &  137.06 &  0.01  &  1.51    &  4482.0  &  -0.77 &  37.33     \\
  29     &  04:00:58.70 & +13:52:26.4 &  136.82 &  0.79  &  0.68    &  3910.0  &  -1.02 &  36.89     \\
  30     &  04:00:58.73 & +13:52:26.8 &  136.82 &  0.79  &  0.63    &  3932.0  &  -0.93 &  35.94     \\
  31     &  04:03:19.70 & +14:56:39.1 &  137.81 &  0.09  &  0.26    &  3934.0  &  -1.08 &  27.64     \\
  32     &  04:59:05.93 & +21:28:01.6 &  152.56 &  -0.28 &  1.25    &  4263.0  &  -0.62 &  18.15     \\
  33     &  04:59:06.00 & +20:30:37.1 &  152.2  &  0.6   &  1.52    &  4426.0  &  -0.74 &  23.61     \\
  34     &  05:00:29.88 & +21:20:42.4 &  152.82 &  -0.05 &  0.35    &  3923.0  &  -1.22 &  35.07     \\
  35     &  05:00:29.88 & +21:20:42.4 &  152.82 &  -0.05 &  0.73    &  3977.0  &  -1.01 &  33.35     \\
  36     &  05:02:03.98 & +21:04:52.0 &  153.05 &  0.34  &  1.29    &  4547.0  &  0.5   &  20.47     \\
  37     &  06:59:45.96 & +21:46:16.7 &  179.6  &  7.27  &  2.28    &  4353.0  &  -0.52 &  25.18     \\
  38     &  07:00:27.79 & +21:46:32.5 &  179.76 &  7.29  &  0.81    &  4421.0  &  -1.23 &  28.01     \\
  39     &  07:00:27.79 & +21:46:32.5 &  179.76 &  7.29  &  1.86    &  4836.0  &  -0.86 &  45.42     \\
  40     &  07:03:33.91 & +21:34:03.7 &  180.44 &  7.61  &  1.86    &  4797.0  &  -1.28 &  49.39     \\
  41     &  07:59:42.55 & +32:51:33.5 &  193.94 &  -2.43 &  1.37    &  4702.0  &  -1.65 &  24.28     \\
  42     &  10:00:04.22 & +18:18:12.2 &  222.36 &  9.25  &  2.11    &  4527.0  &  0.28  &  36.88     \\
  43     &  10:00:04.22 & +18:18:12.2 &  222.36 &  9.25  &  2.25    &  4571.0  &  0.24  &  34.09     \\
  44     &  10:11:48.29 & -01:37:32.5 &  230.65 &  27.81 &  2.1     &  3894.0  &  0.5   &  16.61     \\
  45     &  10:11:52.85 & -01:44:28.3 &  230.71 &  27.92 &  0.84    &  3942.0  &  0.39  &  22.06     \\
  46     &  10:11:52.85 & -01:44:28.3 &  230.71 &  27.92 &  2.17    &  3997.0  &  0.5   &  21.46     \\
  47     &  10:12:26.21 & -01:38:30.1 &  230.83 &  27.78 &  1.86    &  4117.0  &  -1.07 &  17.5      \\
  48     &  10:12:41.59 & -01:32:51.7 &  230.87 &  27.68 &  1.14    &  3974.0  &  -1.72 &  23.26     \\
  49     &  10:12:41.59 & -01:32:51.7 &  230.87 &  27.68 &  1.98    &  4262.0  &  -0.89 &  23.36     \\
  50     &  10:12:41.83 & -01:45:27.4 &  230.93 &  27.88 &  1.87    &  4720.0  &  -1.11 &  37.92     \\
  51     &  10:13:20.74 & -01:31:59.2 &  231.04 &  27.62 &  2.09    &  3509.0  &  -0.61 &  19.4      \\
  52     &  10:13:22.94 & -01:22:27.1 &  231.0  &  27.46 &  1.64    &  4035.0  &  -1.18 &  20.94     \\
  53     &  10:13:34.68 & -01:01:59.2 &  230.94 &  27.12 &  1.94    &  4175.0  &  -0.65 &  21.71     \\
  54     &  10:13:35.30 & -01:17:54.6 &  231.03 &  27.37 &  1.43    &  4497.0  &  -1.11 &  24.19     \\
  55     &  10:13:55.10 & -01:17:03.8 &  231.11 &  27.34 &  1.04    &  4250.0  &  -1.34 &  26.38     \\
  56     &  10:13:55.10 & -01:17:03.8 &  231.11 &  27.34 &  2.21    &  4685.0  &  -1.24 &  19.92     \\
  57     &  10:14:04.30 & -01:13:22.4 &  231.14 &  27.27 &  1.58    &  4080.0  &  -1.38 &  18.51     \\
  58     &  10:14:04.68 & -01:23:46.7 &  231.19 &  27.43 &  1.77    &  4179.0  &  -0.87 &  18.15     \\
  59     &  10:14:39.86 & -01:10:52.7 &  231.28 &  27.18 &  1.05    &  4321.0  &  -1.0  &  19.66     \\
  60     &  10:15:52.08 & -01:33:31.0 &  231.73 &  27.46 &  1.18    &  4113.0  &  -1.29 &  24.13     \\
  61     &  10:58:20.90 & +15:34:44.8 &  236.66 &  7.91  &  1.18    &  4243.0  &  -0.7  &  29.94     \\
  62     &  10:58:20.90 & +15:34:44.8 &  236.66 &  7.91  &  1.91    &  4470.0  &  -0.61 &  37.86     \\
  63     &  11:01:45.26 & +16:00:08.3 &  237.29 &  7.23  &  1.32    &  4604.0  &  -1.11 &  33.88     \\
  64     &  11:02:19.32 & +15:31:30.0 &  237.58 &  7.64  &  1.27    &  3918.0  &  -0.53 &  40.89     \\
  65     &  11:02:19.34 & +15:31:30.4 &  237.58 &  7.64  &  1.34    &  3959.0  &  -0.51 &  32.07     \\
  66     &  11:02:27.36 & +15:35:13.6 &  237.59 &  7.57  &  1.66    &  4833.0  &  -1.48 &  43.44     \\
  67     &  12:25:28.20 & +24:20:53.2 &  251.97 &  -8.25 &  1.66    &  3677.0  &  0.01  &  15.43     \\
  68     &  12:25:35.88 & +23:18:00.0 &  252.48 &  -7.33 &  2.14    &  3559.0  &  0.39  &  26.38     \\
  69     &  12:26:20.54 & +24:31:34.3 &  252.07 &  -8.5  &  1.64    &  4756.0  &  -0.85 &  16.98     \\
  70     &  12:27:24.17 & +23:16:01.9 &  252.87 &  -7.49 &  1.63    &  4082.0  &  -0.48 &  19.86     \\
  71     &  12:29:41.66 & +24:04:06.6 &  252.97 &  -8.44 &  1.09    &  4347.0  &  -1.27 &  26.03     \\
  72     &  12:29:41.66 & +24:04:06.6 &  252.97 &  -8.44 &  0.96    &  4311.0  &  -1.49 &  27.65     \\
  73     &  14:12:45.12 & -00:46:50.5 &  287.07 &  0.91  &  2.1     &  3893.0  &  0.5   &  19.36     \\
  74     &  14:14:29.23 & -00:31:49.4 &  287.31 &  0.47  &  1.57    &  4299.0  &  0.03  &  25.72     \\
  75     &  14:16:22.44 & -01:04:48.0 &  288.0  &  0.7   &  0.42    &  3987.0  &  -1.38 &  39.15     \\
  76     &  14:16:34.99 & -00:21:00.4 &  287.68 &  0.05  &  1.98    &  3640.0  &  0.43  &  24.9      \\
  77     &  14:16:36.60 & -00:10:30.4 &  287.59 &  -0.11 &  2.03    &  4592.0  &  -0.01 &  20.38     \\
  78     &  14:59:10.44 & -06:47:11.4 &  300.09 &  0.26  &  2.17    &  3688.0  &  0.38  &  35.51     \\
  79     &  14:59:58.13 & -07:10:29.3 &  300.46 &  0.5   &  2.21    &  4801.0  &  -0.3  &  24.55     \\
  80     &  15:18:34.13 & +00:17:22.9 &  300.8  &  -8.29 &  0.95    &  3857.0  &  -0.72 &  33.39     \\
  81     &  15:59:00.53 & -14:46:18.8 &  316.8  &  0.23  &  1.37    &  4225.0  &  -0.62 &  22.78     \\
  82     &  15:59:31.73 & -14:18:20.5 &  316.7  &  -0.25 &  0.98    &  3791.0  &  -0.69 &  28.23     \\
  83     &  16:00:05.66 & -14:39:53.3 &  316.99 &  0.01  &  0.97    &  4060.0  &  -0.9  &  27.88     \\
  84     &  16:00:46.25 & -10:08:09.2 &  315.08 &  -4.1  &  1.99    &  4315.0  &  0.07  &  16.32     \\
  85     &  16:01:23.09 & -14:28:50.5 &  317.18 &  -0.3  &  1.92    &  4214.0  &  -0.23 &  30.47     \\
  86     &  17:00:54.10 & -19:57:31.3 &  332.38 &  -1.27 &  1.56    &  4211.0  &  -0.7  &  32.8      \\
  87     &  19:57:50.11 & -33:16:26.0 &  13.56  &  2.86  &  1.32    &  4319.0  &  -0.66 &  23.72     \\
  88     &  19:58:16.32 & -32:46:55.6 &  13.64  &  2.37  &  1.8     &  4499.0  &  -0.2  &  25.2      \\
  89     &  19:58:17.09 & -33:16:32.5 &  13.66  &  2.86  &  2.07    &  4177.0  &  -0.33 &  15.33     \\
  90     &  19:58:39.50 & -32:59:30.1 &  13.72  &  2.57  &  1.84    &  4362.0  &  -0.67 &  17.45     \\
  91     &  19:58:50.18 & -33:05:16.1 &  13.77  &  2.67  &  1.39    &  4225.0  &  -1.23 &  16.2      \\
  92     &  19:59:02.09 & -33:25:51.6 &  13.82  &  3.01  &  1.95    &  4646.0  &  -0.98 &  52.36     \\
  93     &  19:59:48.53 & -33:20:13.6 &  13.98  &  2.91  &  1.67    &  4008.0  &  -0.31 &  23.75     \\
  94     &  20:00:08.81 & -32:42:50.8 &  14.03  &  2.28  &  1.8     &  4370.0  &  -0.68 &  33.01     \\
  95     &  20:00:34.25 & -32:56:00.2 &  14.12  &  2.5   &  1.95    &  4753.0  &  -0.63 &  29.79     \\
  96     &  20:00:48.65 & -33:19:37.2 &  14.19  &  2.89  &  2.25    &  4723.0  &  -1.03 &  30.25     \\
  97     &  20:01:06.31 & -32:41:03.8 &  14.23  &  2.25  &  1.89    &  4264.0  &  -0.6  &  27.7      \\
  98     &  20:01:08.74 & -32:39:25.6 &  14.24  &  2.22  &  1.61    &  4241.0  &  -0.45 &  20.47     \\
  99     &  20:01:12.02 & -32:50:30.8 &  14.25  &  2.41  &  2.16    &  4248.0  &  0.02  &  16.54     \\
  100    &  20:01:20.64 & -32:36:18.7 &  14.28  &  2.17  &  2.18    &  4363.0  &  -0.5  &  31.93     \\
  101    &  20:01:20.81 & -33:20:58.9 &  14.3   &  2.91  &  1.89    &  4586.0  &  -0.82 &  28.07     \\
  102    &  20:02:09.79 & -33:13:50.9 &  14.47  &  2.79  &  1.1     &  4368.0  &  -1.13 &  38.56     \\
  103    &  22:00:16.66 & -30:08:30.5 &  39.72  &  2.29  &  0.81    &  4300.0  &  -1.25 &  40.51     \\
  104    &  22:00:16.66 & -30:08:30.5 &  39.72  &  2.29  &  1.08    &  4299.0  &  -1.15 &  52.7      \\
  105    &  22:00:22.90 & -30:07:50.9 &  39.74  &  2.28  &  1.1     &  4513.0  &  -1.47 &  38.94     \\
  106    &  22:00:22.90 & -30:07:50.9 &  39.74  &  2.28  &  1.03    &  4295.0  &  -1.47 &  44.52     \\
  
\hline
\end{longtable*}
}

\acknowledgments

 The authors would like to thank N. Tothill and M. Filipovic of UWS for discussions regarding this article. We thank the reviewer for valuable comments and changes which have increased the readability of this paper. We would additionally like to thank D. Law, and S. R Majewski for useful discussions on the Sagittarius dwarf, its models and its general behaviour. RRL acknowledges financial support from FONDECYT, project No. 3130403. AK acknowledges the Deutsche Forschungsgemeinschaft for funding. GFL thanks the ARC for support through Discovery Project DP110100678. GFL also gratefully acknowledges financial support through his ARC Future Fellowship (FT100100268). A. R. Conn thanks the University of Sydney for funding via the Laffan Research Fellowship. DBZ acknowledges the support of the ARC in the form of Future Fellowship FT110100743. SLM acknowledges the support of the Australian Research Council through DECRA Fellowship DE140100598. This work was partly supported by the European Union FP7 programme through ERC grant number 320360.

{\it Facilities:} \facility{AAT}

\bibliographystyle{apj.bst}
\bibliography{sgr_2012.bib}

\end{document}